\documentclass[12pt]{article}
\usepackage{scicite}
\usepackage{times}
\usepackage{amsfonts}
\usepackage{amssymb}
\usepackage{amsmath}
\usepackage{graphicx} 

\usepackage{xcolor}
\usepackage[normalem]{ulem}

\newenvironment{sciabstract}{%
\begin{quote} \bf}
{\end{quote}}

\title{Surprising simplicity in the modeling of dynamic granular intrusion} 

\author
{Shashank Agarwal,$^{1\dagger}$ Andras Karsai,$^{2\dagger}$ Daniel I Goldman,$^{2}$ Ken Kamrin$^{1\ast}$\\
\\
\normalsize{$^{1}$Department of Mechanical Engineering, Massachusetts Institute of Technology, Cambridge}\\
\normalsize{$^{2}$Department of Physics, Georgia Institute of Technology, Atlanta}\\
\\
\normalsize{$^\ast$To whom correspondence should be addressed; E-mail:  kkamrin@mit.edu.} \\
\normalsize{$^\dagger$These authors contributed equally to this work.}
}
\date{}

\begin{document} 
\baselineskip24pt
\maketitle 

\begin{sciabstract}
Granular intrusions, such as dynamic impact or wheel locomotion, are complex multiphase phenomena where the grains exhibit solid-like and fluid-like characteristics together with an ejected gas-like phase. Despite decades of modeling efforts, a unified description of the physics in such intrusions is as yet unknown. Here we show that a continuum model based on the simple notions of frictional flow and tension-free separation describes complex granular intrusions near free surfaces. This model captures dynamics in a variety of experiments including wheel locomotion, plate intrusions, and running legged robots. The model reveals that three effects (a static contribution and two dynamic ones) primarily give rise to intrusion forces in such scenarios. Identification of these effects enables development of a further reduced-order technique (Dynamic Resistive Force Theory) for rapid modeling of granular locomotion of arbitrarily shaped intruders. \textcolor{black}{The continuum-motivated strategy we propose for identifying physical mechanisms and corresponding reduced-order relations has potential use for a variety of other materials.}

\end{sciabstract}


\section*{Introduction}
Intrusions into granular media (GM) can create complex flow and force responses, where the media can exhibit both solid-like and fluid-like characteristics. GM \textcolor{black}{deforms} elastically under stress like a solid, but begins to flow like a fluid once a friction-based yield criterion is met. Large variations in the GM's stress, momentum, and volume fraction in different regions often result in  complicated system dynamics exhibiting multiphase characteristics \cite{van2017impact,lohse2004impact}. The flow complexity also makes interpreting resistive forces non-trivial if the intruder re-interacts with the deformed region \cite{Mazouchova2013}, as the GM now has a new inhomogeneous state near the surface. The coupled system of intruder and media becomes challenging to model; the media's inhomogeneous flow and multiphase nature often restricts modeling to discrete particle methods that track the individual grains, unlike fluids that can be solved with the Navier-Stokes equations. \par
\begin{figure}[h!]
\centering
\includegraphics[trim = 0mm 152mm 140mm 0mm, clip, width=0.6\textwidth] {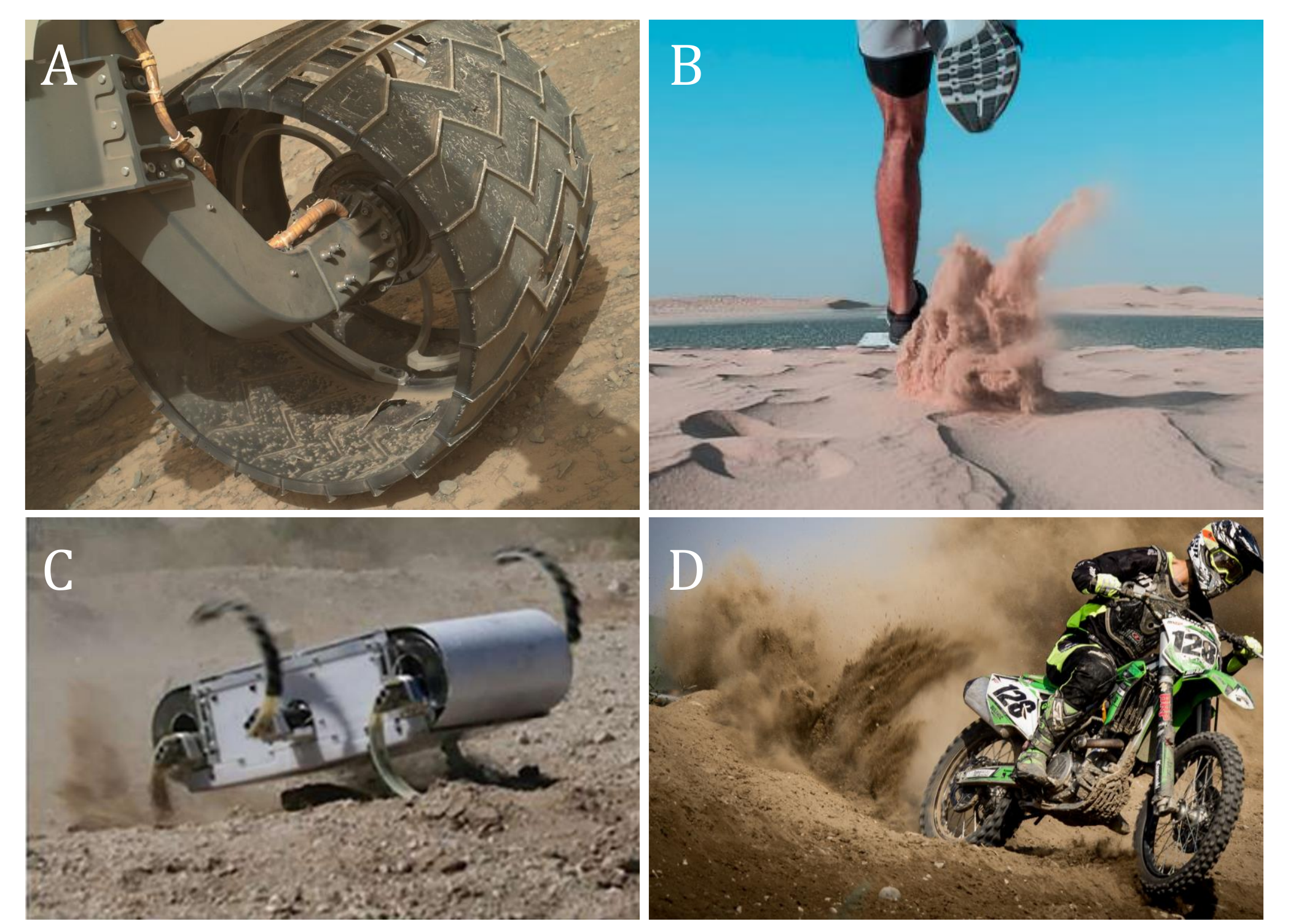}
\caption{\textbf{Examples of locomotion on granular surfaces at various speeds:} \emph{(A)} Wheel of the Curiosity Mars rover (Diameter $\sim$ $50$ cm) \cite{arvidson2017relating} \emph{(B)} running human \cite{img2_alt}, \emph{(C)} RHex C-legged robot (C-leg limb length $\sim$ $18$ cm) \cite{johnson2011autonomous}, and \emph{(D)} a racing dirt bike (Diameter $\sim$ $50$ cm)\cite{img4_alt}. [Photo credits: \emph{(A)} MAHLI imager Curiosity, NASA; \emph{(B)} A. Singh, www.pexels.com; \emph{(C)}  G. C. Haynes, A. M. Johnson, and D.
E. Koditschek, University of Pennsylvania; \emph{(D)} Daniel, www.pexels.com] .}
\label{fig:1}
\end{figure}

A common granular intrusion involves a rigid or flexible solid penetrating into GM and using the resistive force to propel itself into a state of locomotion (see Figure \ref{fig:1}). If a body \textcolor{black}{slowly intrudes into GM}, granular stress arises independent of the intrusion rate, and the resistive force on the intruding body remains in the quasistatic limit \cite{jop2006constitutive,midi2004dense,gravish2014plow}. However, various intrusion scenarios can arise which deform the media rapidly enough that the net force response, and hence the locomotive behavior, is affected. 
Examples of such intrusions include ballistics, meteor impacts, rapid locomotion, and many industrial processes \cite{uehara2003low,melosh1989impact,zhao2015granular,van2017impact,joubaud2014forces}. 

Rigid wheel locomotion is an exemplar of a system that combines these effects, exhibiting multiphase granular behavior, complex grain-surface interactions, and re-interaction with deformed media.  Rigid wheels like those found in planetary rovers \cite{Shrivastavaeaba3499} continuously shear and sometimes violently deform the local GM \cite{wong1967rigidwheels} to locomote in loosely consolidated terrain. These intrusions, particularly in high-velocity cases, cause the substrate material to behavior to deviate significantly from its quasistatic response, driven by potentially non-trivial surface interactions with the wheel. Thus, we first focus on rigid wheel locomotion as a diagnostic scenario of complex intrusion, which includes a wide array of nontrivial effects.  \par

We propose a continuum framework for intrusion based on a frictional-plastic yield condition and free separation. We implement it numerically using the Material Point Method (MPM) \textcolor{black}{(more details in the Material and Methods section)}. \textcolor{black}{Our intrusion analysis begins with a focus on driven circular wheels with grousers \textcolor{black}{ --- grousers are finite-sized radial protrusions along the wheel circumference, which facilitate traction (See figure \ref{fig:2}A, \ref{fig:4}A, Movie S3 and Table S1 for more details).   Grousered wheels are commonly used in granular locomotion applications in soft terrain \cite{ding2011experimental,agarwal2019modeling,suzuki2019study,Shrivastavaeaba3499,wong1967rigidwheels}.}} Alongside scenarios of slow and rapid wheeled locomotion, two additional families of test cases, submerged lateral plate intrusion and ``four-flap runners,'' are simulated and compared to know results in the literature to verify the model's ability to capture dynamics of complex granular intrusions. Interestingly, our proposed continuum model captures the non-trivial \emph{rate-dependent phenomena} exhibited in complex intrusions even though its constitutive equations are \emph{rate-independent}.  Our work shows how a \textcolor{black}{single} continuum interpretation of GM can represent multiple intrusion scenarios by implicitly reconciling various inertial effects. 

 \textcolor{black}{Importantly, we also obtain a global-level physical understanding of intrusion dynamics by analyzing plasticity solutions, which guides the development of a reduced-order model for intrusion that we call the Dynamic Resistive Force Theory (DRFT).  
 We show that DRFT accurately models all the considered granular intrusion cases.} By combining existing literature, continuum modeling, and experimental verification, we identify the relevant physics that go into DRFT and its interpretation as corrections to an existing quasistatic RFT model \cite{li2013terradynamics,zhang2014effectiveness,agarwal2019modeling} for slow intrusion. \textcolor{black}{Key effects that generate rate-dependent behaviors} are identified, and, once incorporated, DRFT allows rapid calculation of the expected resistive forces in GM. 

\section*{Results and Discussion}
\subsection*{Wheel Locomotion Experiments}
\begin{figure*}[ht!]
\centering
\includegraphics[trim = 0mm 163mm 92mm 0mm, clip, width=1.0 \linewidth] {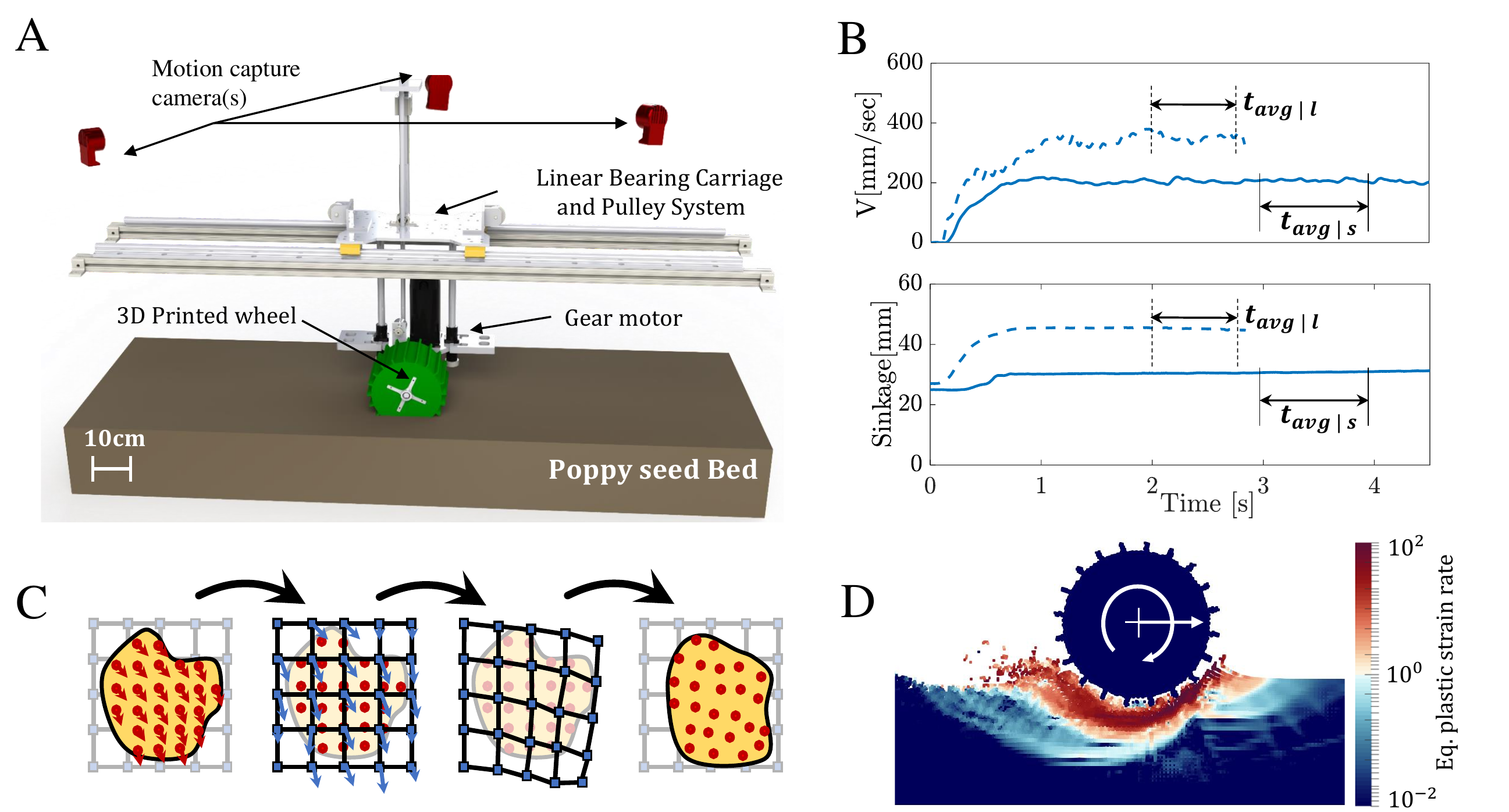}
\caption{\textcolor{black}{\textbf{Setup for rigid-wheel experiments and continuum simulations:}} \emph{(A)} CAD model of experimental setup.  \emph{(B)} Sample experimental time series data for translation velocity (Top) and sinkage (Bottom) at low $\omega$ (20 RPM, solid lines) and high $\omega$ (50 RPM, dotted lines) respectively. $t_{avg|s}$ and $t_{avg|l}$ show the time windows used for avergaing low and high $\omega$ data, respectively. \emph{(C)} Schematic representation of explicit-time integration in a Material Point Method (MPM) step, whereby a background grid assists in integrating the motion on a set of continuum material points. Solid circles (red) are material points (Lagrangian tracers) and squares (blue) are the nodes of the background mesh. \emph{(D)} A sample continuum simulation using MPM. The field being plotted is the equivalent plastic strain-rate.} 
\label{fig:2}
\end{figure*}

Figure \ref{fig:2}A shows a CAD model of the \textcolor{black}{laboratory setup used for performing wheeled locomotion experiments} 
in this study and Figure \ref{fig:2}B indicates \textcolor{black}{our} data collection methodology. More details of the experimental setup are provided in the materials and methods sections (and Movie S3 of the Supplemental Information). \par

Figure \ref{fig:3}A and B shows the trends of steady-state translation velocity and sinkage (resp.) with increasing angular velocity for a grousered wheel's free locomotion. Experiments indicate the emergence of a rate-dependent effect in wheel locomotion; an increase in slipping, accompanied by an increase in the sinkage of the wheels, breaks the linear trend {in velocity vs $\omega$} seen in the quasistatic domain of $\omega <30$ RPM (corresponding to $\omega/\omega_o < 0.46$ in Fig \ref{fig:3}A). 

\begin{figure*}[ht!]
\centering
\includegraphics[trim = 5mm 165mm 90mm 0mm, clip, width=1.0 \linewidth] {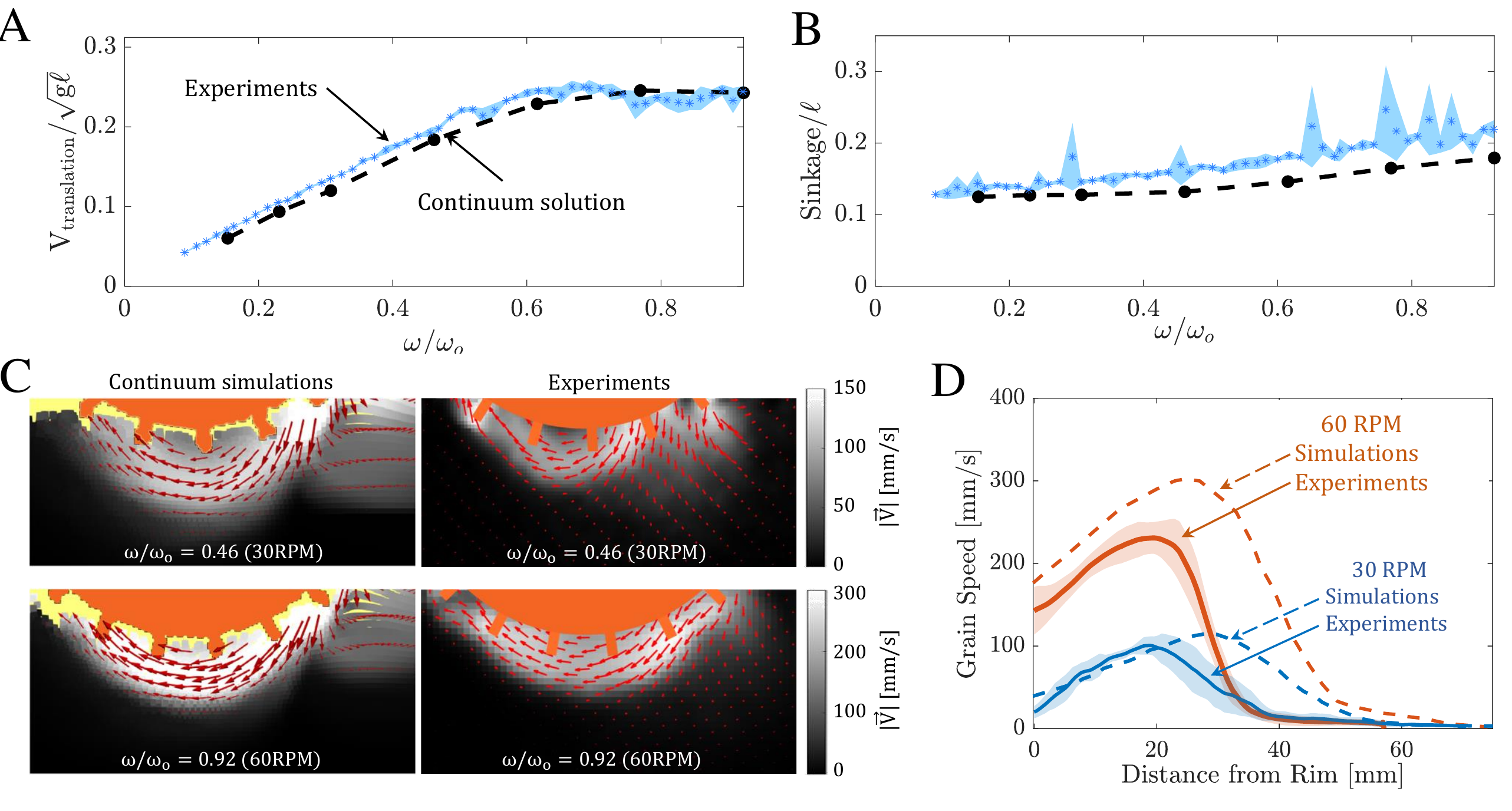}
\caption{\textcolor{black}{\textbf{Comparison of wheel locomotion experiments and continuum simulations:}} Variation of \emph{(A)} translation velocity, and \emph{(B)} sinkage from experiments (blue), and continuum modeling solutions (black). The results are non-dimensionalized using a characteristic system velocity $(g\ell)^{0.5}$ ($= 1440$ mm/s) for translation velocity; a characteristic system length $\ell$ ($=212$ mm) for sinkage; and a characteristic angular velocity, $\omega_o$ ($ =(g/\ell)^{0.5} = 65$ RPM) for angular velocity, where $g$ represents the gravity and $\ell$ represents the wheel's outer diameter. \emph{(C)} Granular flow field velocities obtained from continuum modeling and experiments (PIV) for slow (30 RPM, top) and fast (60 RPM, bottom) wheel locomotion. Data are averaged over an effective rotation of 0.1 rad (for PIV), with the orange regions representing the mean position of the wheel. See Movie S4, S5, and S6 for more details. \textcolor{black}{\emph{(D)} Plots showing variation of grain velocity from continuum simulations and PIV experiments along the radial direction directly below the center of the wheel; note some wall friction from the plexiglass plate exists in the experiment but not in the continuum solution. Key structural features of the flow under the wheel agree between the experiments and model in \emph{(C)} and \emph{(D)}}} 
\label{fig:3}
\end{figure*}

\subsection*{Continuum modeling analyses of granular intrusion}
We consider a simple granular continuum model, which \textcolor{black}{captured} intruder dynamics in previous studies in the slow, quasistatic regimes  \cite{agarwal2019modeling,askari2016intrusion}. Poppy seeds (PS), the granular media used in this study, are modeled as a granular continuum with a \textcolor{black}{Drucker}-Prager (rate-independent friction-based) yield criterion, incompressible plastic shear behavior, and a criterion that the material separates into a stress-free media when brought below a critical density. This rheology can be defined by two simultaneous constraints shown below, describing the material's separation behavior and shear yield condition:
\begin{align}
&\textrm{Free separation:}&&(\rho-\rho_c)P=0 \quad \textrm{and} \quad P\ge 0 \quad \textrm{and} \quad \rho\le\rho_c  \label{eq:4} \\
&\textrm{Frictional yielding:}&&\dot{\gamma}({\tau}-\mu_s P)=0 \quad \textrm{and} \quad \dot{\gamma}\ge 0 \quad \textrm{and} \quad {\tau}\leq\mu_{s} \textrm{\textcolor{black}{$P$}} \label{eq:5} 
\end{align}
for $i,j=1,2,3$. We define
${\sigma}'_{ij}$= ${\sigma}_{ij}$ + $P\delta_{ij}$ as the deviatoric part of the Cauchy stress tensor, 
$P = -{\sigma}_{ii}/3$ as the hydrostatic pressure, 
${\tau}$  =  $\sqrt{\sigma'_{ij}\sigma'_{ij}/2}$ as the equivalent shear stress, $\mu_s$ as the bulk friction coefficient, and 
$\rho_c $ as the critical close-packed granular density. The (plastic) flow rate tensor is $D_{ij}=(\partial_iv_j+\partial_j v_i)/2$ and $\dot{\gamma}=\sqrt{2 D_{ij}D_{ij}}$ is the equivalent shear rate. When shearing plastically, the stress and flow-rate are presumed to align (e.g. $\sigma_{ij}'/2{\tau}=D_{ij}/\dot{\gamma}$). The model evolves the flow by solving the momentum balance equations, $\partial_{j}\sigma_{ij}+\rho g_i=\rho \dot{v}_i$.  Below the yield criterion, the grains act like a linear-elastic solid, \textcolor{black}{so that our model is in fact elastic-plastic in the dense regime}. We assume a constant surface friction coefficient describes the interaction of the granular continuum with solid-body  surfaces. The input material properties in {Table S1 (in the Supplementary Information)} are used. 

We use the Material Point Method (MPM) algorithm described in Dunatunga and Kamrin \cite{dunatunga2015continuum,dunatunga2017continuum} to implement these constitutive equations assuming 2D plane-strain motion. {A schematic representation of an explicit time-integration MPM step is shown in Figure \ref{fig:2}C; the material points carry the continuum data and are moved each step with the help of a background grid}. {Figure \ref{fig:2}D shows a sample wheel-locomotion using MPM, plotting the variation of equivalent plastic shear-rate in the system. }


The trends of steady-state translation velocity and sinkage with varying $\omega$ obtained using continuum modeling are plotted in Figure \ref{fig:3}A and B. Continuum modeling successfully captures the experimental trends for wheel locomotion; in particular, the model captures the plateau in the normalized $v-\omega$ curve at the correct rotation speed and correctly predicts increased sinkage with rotation rate. 
To check robustness of the results, we also applied small changes to the initial state of the experimental and simulated systems --- including  minor variations in initial wheel depth, initial wheel velocity, and ramp-rate of the wheel --- and observed that the steady-state results in both systems were insensitive to these variations.

\par
To further validate the model predictions, we have conducted experiments to visualize subsurface flow fields and compared them to the model. The experiments place the wheel adjacent to a clear plexiglass plate so a camera can capture the underlying grain motion with Particle Image Velocimetry (PIV) \cite{gravish2014plow}. Velocity fields in grains for 30 and 60 RPM cases from continuum modeling and experimental PIV analysis are plotted in Figure \ref{fig:3}C and D. \textcolor{black}{There is some wall drag from the plexiglass plate that likely causes the granular flows in the experiment to be overall slower than the model, however, the key structural features of the flow under the wheel agree between the experiments and model.} Importantly, both show a zone of material ahead of the wheel being pushed forward and a wide zone under and behind the wheel being pushed to the rear. The rear flow zone also grows with increasing $\omega$ due to higher flow entrainment and material movement at higher $\omega$.\par 
\par

\subsection*{\textcolor{black}{Towards reduced-order models}}
\textcolor{black}{
A major benefit in identifying an accurate continuum model for a system is the possibility of using it to extract global-scale simplifications of the system's dynamics that can be used to  develop further-reduced models.} For example, in previous work on slow quasistatic intrusion, Askari et al. \cite{askari2016intrusion} found a connection between frictional yielding and a reduced-order intrusion force model called granular Resistive Force Theory (RFT) \cite{li2013terradynamics}. \textcolor{black}{ The success of the present continuum model for slow and rapid locomotion in wheels (and the other intrusion scenarios in this study) motivates us to ask whether an RFT-like reduced-order model for complex and rapid intrusions exists, and if it might be derivable based on phenomena observed within the continuum model.} We begin by first defining the quasistatic form of RFT and evaluating its predictions for wheeled locomotion dynamics.

RFT is an empirical methodology that has been successful in estimating the force response for arbitrarily-shaped intruding geometries in the quasistatic limit, permitting direct simulation of locomotion in granular volumes \cite{li2013terradynamics,zhang2014effectiveness,agarwal2019modeling}. 
\textcolor{black}{RFT assumes the stress on a small surface element of an intruder follows a \emph{localized} formula in which the surface stress depends only on the motion, location, and orientation of that element \cite{maladen2009undulatory}.} {\textcolor{black}{This local formula decouples the stress response among the suface elements of an intruder, thereby permitting RFT to predict intrusion forces with calculations that run in near real time.}}
In a coordinate system where $z$ points positive upward with granular free surface at $z=0$, and $x$ is a chosen horizontal axis perpendicular to $z$, RFT presumes the force-per-area vector (or traction) $\mathbf{t}$, on each surface element  can be written as  $\mathbf{t}=\boldsymbol{\alpha}(\beta, \gamma)\, H(-z)|z|$, dependent on the element's orientation angle ($\beta$), velocity angle ($\gamma$), and vertical depth from the free surface ($|z|$),  with $H$ being the Heaviside function. The empirical traction-per-depth vector  $\boldsymbol{\alpha}(\beta, \gamma)=(\alpha_x(\beta, \gamma),\alpha_z(\beta, \gamma))$ is measured with small plate intrusion experiments which vary $\beta$ and $\gamma$. By summing these locally-defined tractions, RFT predicts the net resistive force and moment on the entire intruder surface $S$. For example, RFT gives the following intrusion force formula:
\begin{equation}
\mathbf{F} = \int_S \boldsymbol{\alpha} (\beta , \gamma)\,  H(-z) |z|\, \text{dA}.
\label{eq:RFTformula1}
\end{equation}

\begin{figure*}[ht!]
\centering
\includegraphics[trim = 5mm 174mm 45mm 0mm, clip, width=1.0\linewidth] {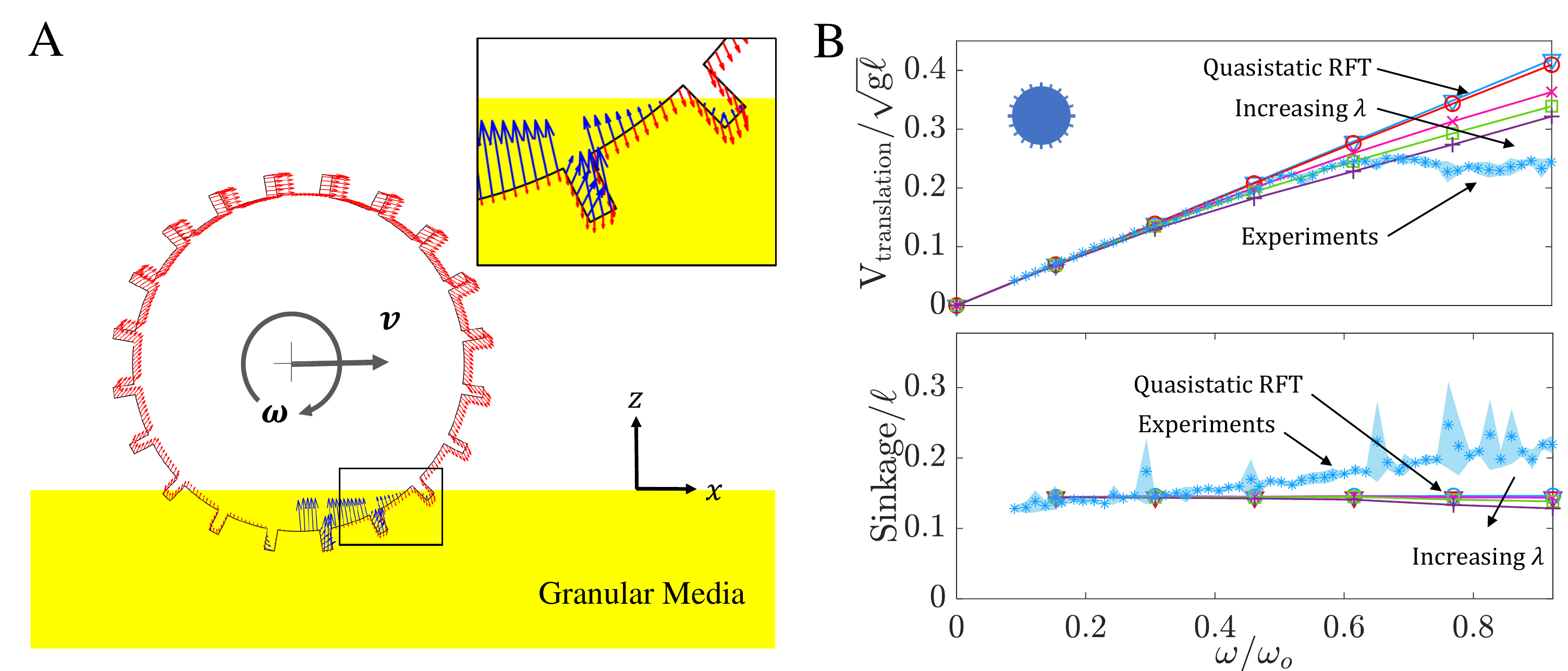}
\caption{\textbf{Experiments vs RFT:} \emph{(A)} Snapshot of a quasistatic RFT simulation, used for studying grousered wheel locomotion. Direction and magnitude (normalized) of the velocity and resistive stress are indicated by red and blue arrows, respectively, along surface elements of the wheel boundary.
\emph{(B)} Translational velocity (Top) and sinkage behaviors (Bottom) of the wheel;  experimental mean and $1\sigma$ standard deviation (light blue data) and RFT results with local $\lambda \rho v^2 $ modification (solid lines).   \textcolor{black}{The results in \emph{(B)} are non-dimensionalized as explained in figure \ref{fig:3}.} The direction of increasing $\lambda$ is indicated ($\lambda$=$0,1,25,50,100$). Red solid lines with $\lambda=0$ (in \emph{B}) correspond to quasistatic RFT results.}
\label{fig:4}
\end{figure*}

Figure \ref{fig:4}A and B show \textcolor{black}{the results of applying} quasistatic RFT (solid blue line in Figure \ref{fig:4}A and B) in modeling grousered wheel locomotion. In implementing the RFT model of locomotion, we also use a `leading edge hypothesis' to ensure that resistive forces experienced by the wheel consist of contributions only from surface elements which move `into' the sand, i.e. surfaces whose outward normal ($\boldsymbol{n}$) and velocity ($\boldsymbol{v}$) make a positive inner product ($\boldsymbol{n}\cdot\boldsymbol{v}>0$). Using the established RFT functions $\alpha_{x}$ and $\alpha_z$, for the granular media used in our experiments \textcolor{black}{\cite{li2013terradynamics}}, RFT was applied to the spinning wheel geometry. Figure \ref{fig:4}B shows that while RFT captures the speed vs $\omega$ trends at low $\omega$, at higher $\omega$ it does not predict the wheel locomotion kinematics. RFT predicts a linear relation between the angular and translation velocities, which matches the experiments' dynamics at low-speeds, but diverges as $\omega$ increases. The fact that quasistatic RFT predicts the steady speed of a round wheel to always be a constant multiple of the wheel spin can be shown as a consequence of the rate-independence of the RFT traction relation in equation \ref{eq:RFTformula1} (see Section S2 of the Supplementary Information for more details).  \par

{\color{black}
\subsection*{Exploiting the continuum treatment for physical insight}
\textcolor{black}{An} important step in developing a general reduced-order model for high-speed granular intrusion scenarios is to identify the key underlying physics. In granular intrusions, rate effects could arise due to a variety of physical causes. Increased vibrations in the media could fluidize \textcolor{black}{the material} at high speeds and reduce its strength \textcolor{black}{\cite{rubin2006failure}}. Increasing velocities could decrease the friction on the wheel/media interface (per a dynamic friction drop), which in turn could decrease the traction on the wheels. Rapid flows may also have significant
 \emph{micro-inertia}, which makes the rheology rate-dependent by causing the stress ratio $\mu\equiv \tau/P$ to depend on shear-rate through the ``inertial number'' $I$, where $I=\dot{\gamma}\sqrt{d^2\rho_s/P}$, where $\dot{\gamma}$ is the shear-rate, $d$ the mean grain diameter, $\rho_s$ the solid particle density, and $P$ the local pressure \cite{jop2006constitutive,midi2004dense}.
Moreover, conventional \emph{macro-inertia} (i.e. the $\rho \dot{v}_i$ term in the momentum balance) adds inertial body forces that could alter the flow of the media and its resistance against the intruder.} 

Predicting the dominating rate effect(s) is difficult using experiments alone. In this regard, our continuum modeling approach greatly aids in eliminating non-significant candidates from the possible rate effects above. \textcolor{black}{ The key is to recall that our model implements a \textit{rate-insensitive frictional surface interaction} with no dynamic friction drop on the wheel-sand interface, and a \textit{rate-insensitive constitutive model} with no dependence on the inertial number nor any accounting of material thermalization or fluidization. The model does, however, include macro-inertia in the momentum balance equations.} The fact that the continuum model is successful in capturing the wheel dynamics along with many other granular intrusion scenarios (discussed later), indicates that the observed rate effects can be reconciled solely from macro-inertia ($\rho\dot{v}_i$). \textcolor{black}{ At the same time, the \emph{global} consequences of \emph{local} macro-inertial forces may be subtle and depend upon the particular system and its dynamics.} \par
\textcolor{black}{Based on this insight, along with analysis of the continuum solutions to wheel locomotion and other granular intrusion scenarios from the literature, we now propose and test a more general RFT that  encompasses the domain of slow to rapid intrusions in granular media, which we refer to as ‘Dynamic RFT’ (DRFT).
}

\subsection*{\textcolor{black}{Dynamic RFT}}
DRFT modifies the quasistatic RFT in two ways to account for macro-inertial effects. First, we add a momentum flux contribution, which we term the \emph{dynamic inertial correction}. This term is required for the transfer of momentum to the granular material surrounding the intruder. This term is also in accord with many previous studies on high-speed granular intrusions \cite{katsuragi2007unified,clark2014collisional,umbanhowar2010granular,clark2016steady,clark2012particle,aguilar2016robophysical,bester2017collisional}, and takes the form of an additional rate-dependent force going as velocity-squared. The second modification, which we will show is critical for more complex intrusions, describes the way in which increased bulk inertia can change the free-surface geometry. A change to the free-surface geometry then feeds back on the resistive forces through the depth-dependence of RFT. We denote this modification as the \emph{dynamic structural correction}. Taken together, DRFT imposes the following formula for the traction on a surface element:
\begin{equation}
    \mathbf{t}=\boldsymbol{\alpha}(\beta, \gamma) H(-\tilde{z})|\tilde{z}| -\mathbf{n}\lambda \rho v_n^2
    \label{DynamicRFTsurfacetraction}
\end{equation}
where $|\tilde{z}|$ indicates the \emph{effective depth} of the surface element. That is, $\tilde{z}=z+\delta h$ where $\delta h$ represents the height decrease of the free surface in the zone affecting the traction at $(x,z)$. Recall $\mathbf{n}$ represents the outward normal to the surface element (and $-\mathbf{n}$ the inward), and we define $v_n$ as the normal component of the surface velocity. 
To use DRFT, one must determine the appropriate $\delta h$ for each surface element of the intruder as a function of the intruder motion as well as an appropriate $\lambda$, an $O(1)$ scalar fitting constant. \textcolor{black}{Similar to RFT, DRFT \textcolor{black}{asserts a \textit{localized} formula for the calculation of stresses on intruder sub-surfaces,} and thus allows for near real-time modeling of intruder motion.}  \par

\textcolor{black}{The procedure for modeling intrusions with DRFT begins by discretizing the intruding geometry into subsurfaces of size $\Delta \ell$, and defining the system properties such as intruder weight, gravity, the effective media density $\rho$, and RFT scaling coefficient $\xi$ from calibration. One can then identify the leading edges of the intruder, and use an external measurement to determine $\lambda$ and the effective free surface variation $\delta h$. Scaled continuum simulations, DEM, or experiments (PIV) can be used to measure $\lambda$ and the behavior of $\delta h$; we provide an example of a methodology to infer $\delta h$ using scaling \textcolor{black}{analysis} in the Supplementary Information (Figure S1 and Section S3). Equation \ref{DynamicRFTsurfacetraction} can then \textcolor{black}{be used to} calculate the granular resistive force on the leading edges. With intrusion force fully defined by DRFT, one can iteratively apply a momentum balance in different directions on the intruder geometry to model the intruder motion. 
} \par

\par

\subsubsection*{\textcolor{black}{Understanding the dynamic inertial correction}} 
\textcolor{black}{We take a moment to discuss the two dynamic corrections included in DRFT, beginning with the dynamic inertial correction.}  
Analyis of the momentum balance equations under certain simplifying circumstances (see Supplementary Information, Section S1) allows one to deduce that the transition from a quasistatic flow to a faster flow comes with a resistive force increase that goes as $\rho A v_n^2$, similar to dynamic pressure in a fluid, where $A$ is the intruder area.  Physically, this term represents the reaction force that comes from transferring momentum to the granular media.
 A number of previous studies \cite{katsuragi2007unified,clark2014collisional,umbanhowar2010granular,clark2016steady,clark2012particle,aguilar2016robophysical,bester2017collisional} have modeled the rate-dependence of intrusion force similarly, by adding a term proportional to normal speed squared to a depth-dependent `static' term. 
 Examination of experimental data in  \cite{umbanhowar2010granular,schiebel2019mitigating}, agrees with a rate-dependent force addition of the form $\lambda \rho A v_n^2$ in simple vertical and horizontal intrusions (see Figure S2 and S3, and Movie S1 and S2 of the Supplementary Information), where $\lambda$ is a $O(1)$ scalar fitting constant that accounts for certain approximations in the analysis (see Supplementary Information, Section S1). 

It is natural to ask whether the addition of a velocity-squared term to the quasistatic RFT relation is enough alone to explain the rate-dependence observed in general intrusion scenarios, including wheeled locomotion. We suppose the surface traction is modeled to obey the relation in equation \ref{RFTsurfacetraction} below, and use this relation to re-evaluate the grousered wheeled locomotion problem:
\begin{equation}
    \mathbf{t}=\boldsymbol{\alpha}(\beta, \gamma) H(-z)|z| -\mathbf{n}\lambda \rho v_n^2
    \label{RFTsurfacetraction}
\end{equation}
Figure \ref{fig:4}B shows the results  for various values of $\lambda$. The case of $\lambda=0$ represents the previously discussed quasistatic RFT in these graphs. The introduction of the inertial force term ($\lambda>0$) adds a new force contribution having net force components upward and opposite to the horizontal direction of wheel translation. This upward force results in a decrease of wheel sinkage, opposite to the experimental observation. The magnitude of these extra forces is very small; the pre-factor $\lambda$ was varied from 1 to 100 in an attempt to match the experiments, but this has little effect on the outcome and the trends for both velocity and sinkage cannot be matched (Figure \ref{fig:4}B). It is clear the dynamic inertial correction alone is not sufficient to describe this set of tests.\par

\begin{figure*}[t!]
\centering
\includegraphics[trim = 0mm 128mm 60mm 0mm, clip, width = 1.0\linewidth] {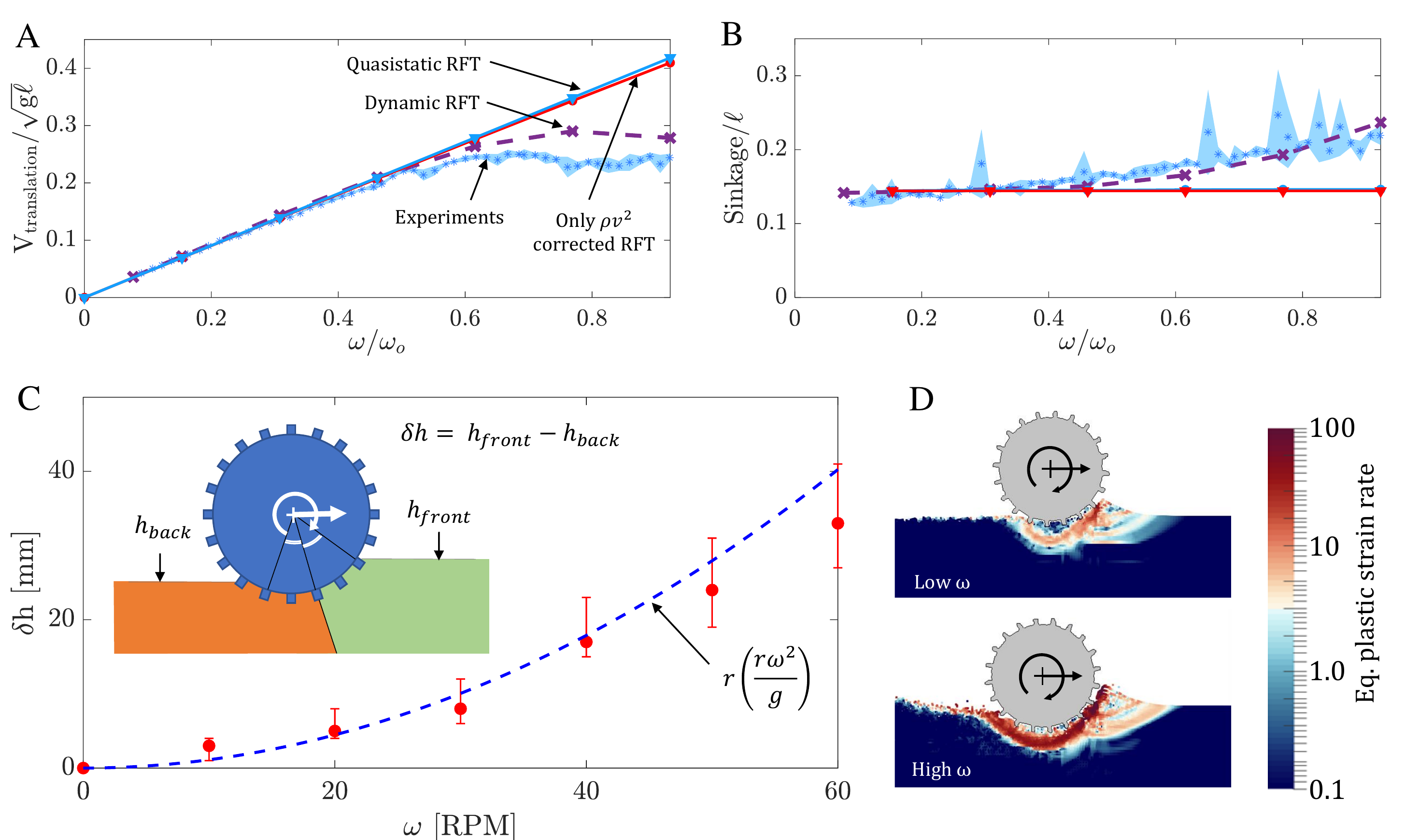}
\caption{\textbf{\textcolor{black}{Dynamic RFT:} } Variation of \emph{(A)} wheel translation velocity and \emph{(B)} sinkage from experiments compared to quasistatic RFT, DRFT, \textcolor{black}{and DRFT without any dynamic structural correction (i.e. having only the $\sim\rho v^2$ correction)}. \emph{(C)} Presumed zones of influence and effective free-surface variation for constructing the dynamic structural correction; $\delta h$ represents the gap between effective front and back free surface positions. MPM data (red circles) and empirical fit (blue dotted line) for $\delta h$.  \emph{(D)} Variation of equivalent plastic strain-rate magnitude obtained using MPM continuum modeling for slow (30 RPM) and high speed (90 RPM) wheel locomotion. See Movie S7 in the Supplementary Information for visualising the variation over time. \textcolor{black}{The results in \emph{(A)} and \emph{(B)} are non-dimensionalized as explained in Figure \ref{fig:3}.} 
}
\label{fig:5}
\end{figure*}

\subsubsection*{\textcolor{black}{Understanding the dynamic structural correction}}
\textcolor{black}{To understand the rationale behind the dynamic structural correction in DRFT, we start by considering the spatial variation of plastic strain-rate magnitudes from continuum modeling simulations for low and high $\omega$ cases shown in Figure \ref{fig:5}D}. The plots make it possible to visualize how different portions of the wheel derive their resistive forces from different zones of the granular media. While the strain-rate profiles change as angular velocities increase, the basic patterns of shearing remain similar. The sheared material reaches the free surface of the granular volume in two zones. Approximately half of the flow originating from the leading edge of the wheel reaches the free surface on the trailing rear face of the wheel. The remaining flow-lines extend to the free surface on the leading front face of the wheel. Importantly, the height of the free-surface on the rear side of the wheel decreases with increasing $\omega$; qualitatively, as $\omega$ grows, the wheel expels material on the rear side. The reduction in rear free surface height suggests a reduction in the pressure head and consequent weakening of the material in the rear shear zone. This is a key observation, which motivates the form of the dynamic structural correction.\par 


Figure \ref{fig:5}C shows the free surface height reduction, $\delta h$, as measured from the continuum model simulations by identifying the lowest point making rear contact with the wheel for which hydrostatic-pressure $\to 0$. Indeed, the faster the wheel spins, the deeper this point descends. Given the paucity of parameters in the continuum model, dimensional analysis is useful; for a given substrate material it suggests the form $\delta h =r\cdot \psi(r\omega^2/g)$ for some function $\psi$.  Surprisingly, we find that $\psi$ is well-approximated by the identity function. The fit of $\delta h=r\, (r\omega^2/g)$ and the continuum modeling results in Figure \ref{fig:5}C show good agreement. Combined with the understanding developed in the previous section, the form of the effective free surface is approximated using a simple partition as shown in Figure \ref{fig:5}C, with the rear zone of the wheel set to have a constant free-surface height reduction $h_{back}$ differing from the initial free-surface height (undisturbed medium height) by a term $\delta h=r\,(r\omega^2/g)$. \textcolor{black}{To select the dividing angle delineating the front- and rear-affected zones of flow, we choose to equally divide the contact zone for driven wheels. Our choice is driven by the simplicity of this division, also observing a similar division of contact zones for representing traction on wheels by Hambleton et al. \cite{Hambleton2009}.} This new model changes the effective free-surface heights only for the surface elements closer to the rear part of the intruding wheel surface. 

By including this effective free-surface height formulation, we now arrive at DRFT, equation \ref{DynamicRFTsurfacetraction}. We implement this DRFT model using the same implicit RFT code framework discussed in the Materials and Methods section, using $\lambda=1$ and $\rho\approx\rho_c=638$ kg/m$^3$. The trends of translation velocity and sinkage with respect to $\omega$ now show good agreement between experiment and DRFT (Figure \ref{fig:5} A and B). \textcolor{black}{We also include, for comparison, what the solution is when only the dynamic inertial correction is used.} While DRFT combines both dynamic corrections, it is clear that the dynamic structural correction dominates the dynamic inertial correction in the case of wheeled locomotion. \textcolor{black}{While we have presumed for simplicity that the division between the two contact zones takes place halfway through the wheel-sand interface, it can be seen in Figure \ref{fig:3}C and \ref{fig:5}D that the division may actually be closer to the front of the wheel. This could explain our slight overprediction of speed for high $\omega$ (Figure \ref{fig:5}A). A second set of grousered wheels tests involving a smaller wheel are included in the Supplementary Material (see Figure S1 and Section S3) and DRFT works equally well without the need to refit the function for $\psi$ used for $\delta h$.}

The agreement with DRFT suggests that the low-to-high slip  transition in wheeled locomotion (where slip$=1-v/r\omega$ for $v$ is the translational velocity, $r$ is the nominal radius, and $\omega$ is the angular velocity of the wheel) occurs largely because faster spinning wheels remove material from behind the wheel, which reduces the pressure in the rear zone, thereby weakening the base of material that would otherwise provide a scaffold off of which the wheel pushes. Updating RFT by accounting for this effect has appropriately captured the dynamics of the complex wheel locomotion scenario in a reduced-order modeling framework.\par

\subsection*{Additional verification studies for the continuum model and DRFT}
\begin{figure*}[ht!]
\centering
\includegraphics[trim = 5mm 180mm 40mm 0mm, clip, width=1.0 \linewidth] {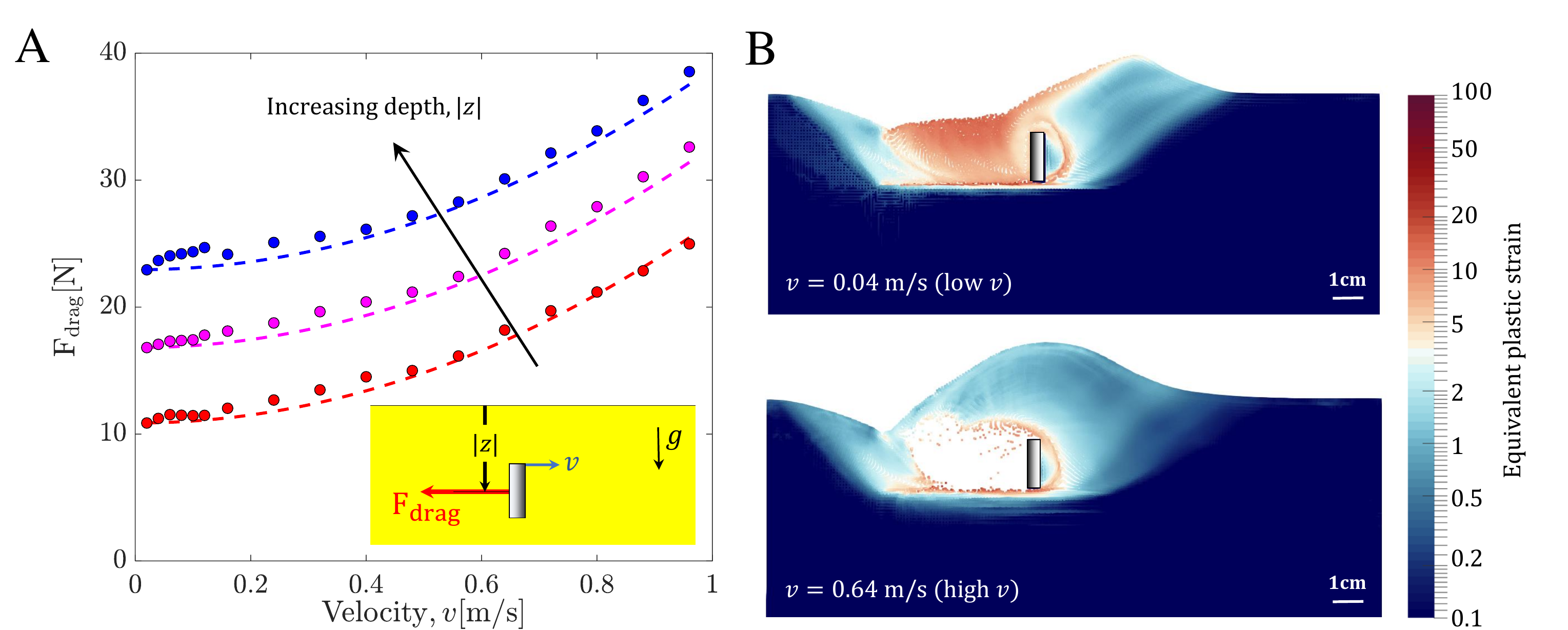}
\caption{\textbf{Modeling slow-to-rapid plate intrusion:} \emph{(A)} Continuum MPM data (colored circles) and ${K|z| + \lambda\rho A v^2}$ fits (dotted lines) for horizontal intrusions at various depths ($|z|$) ([20,30,40]  mm), where $K = 580$ N/m and $\lambda = 1.1$. Variation of equivalent plastic strain rate for \emph{(B, top)} low Velocity (0.04 m/s) and \emph{(B, bottom)} high velocity (0.64 m/s) intrusion cases (at 30  mm depth). See Movie S2 in the Supplementary Information for the video. Simulations are plane-strain.}
\label{fig:6}
\end{figure*}

The wheel tests provide a complex intrusion scenario, and have a dynamic structural correction that is much larger than the inertial correction. To check the robustness of our continuum modeling approach as well as Eq \ref{DynamicRFTsurfacetraction} for DRFT we now examine the converse situation \textcolor{black}{ with two additional sets of simulations --- submerged plate intruders and locomoting runners. We choose these cases based on data from continuum solutions, validations against the literature, and the arguments of the previous section, and expect the dynamic structural correction to be small and the dynamic inertial correction to dominate. Visually, these cases represent two separate classes of intruders. While the dragged plates represent forced motion, the runners represent a class of self-propelling locomotors which may appear similar to the prior studied wheels. Yet, force responses in both cases are dominated by the dynamic inertial correction (more details in following sections) and do not mimic behavior of the grousered wheel. Thus, these distinct cases test the breadth of the modeling capability of DRFT. 
}
\\
\\
\noindent\textit{Submerged horizontal intrusion}: Thin plates submerged in a granular media at various fixed depths (20-40 mm) are dragged horizontally at different speeds \textcolor{black}{using continuum modeling}. The continuum model runs in plane strain, where the plate has a length of $0.016$ m and the effective medium density is 900 kg/m{$^2$}. \textcolor{black}{The chosen density is similar to that of ground coal or marble}. The filled circles in Figure \ref{fig:6}A show the 
\textcolor{black}{observed} drag force variations with 
the drag speed. Experimental studies by Schiebel et al. \cite{schiebel2019mitigating} found the variation of drag forces in such a scenario to follow the trend $K|z|+\lambda \rho A v^2$ (see Figure S1 of Supplementary Information), where $K$ and $\lambda$ are constants, $|z|$ is the depth of the plate below the free surface,  $\rho$ is the effective granular density, $A$ is plate area, and $v$ is horizontal plate velocity. Our continuum modeling also obtains the same trend (Figure \ref{fig:6}A). In the slowest cases ($v\sim0$), we obtain a linear force versus depth relation, $F_{\text{drag}}=K|z|$ for $K=580$ N/m. As speed increases, we find that continuum predictions match the data well at three different depths, for $\lambda = 1.1$. Incidentally, the same value of $\lambda$ also matches the rate dependence observed in the Schiebel et al. \cite{schiebel2019mitigating} experiments for horizontally driven intruders at the free surface.  \par

 A comprehensive understanding of the resultant form of the drag force trends can be obtained by observing continuum modeling results in the context of DRFT. Figure \ref{fig:6}B shows the deformation profiles around the plate at two selected speeds (which differ by about an order of magnitude). The profiles in Figure \ref{fig:6}B of high and low speed intrusion suggest that the intruder tractions arise from pressing the granular material in front of the plate upward and to the right, toward a common free surface height, $h_{\text{front}}$. The rear flow zone, which changes in slow versus high-speed intrusion, is either in the separated phase or newly consolidated as it falls and fills in the gap behind the moving plate. \textcolor{black}{ Likewise, the rear media makes a negligible contribution to the resistive plate force;  
 \textcolor{black}{no part of the} of the rear face of the plate is a `leading edge' satisfying $\boldsymbol{n}\cdot\boldsymbol{v}>0$, so forces approximately vanish there. This is in contrast to the grousered wheel case, where, due to rotation, a significant portion of the back half of the wheel is a leading edge} that can interact non-trivially with media behind the wheel. We thus expect a negligible dynamic structural correction for horizontal plate drag, due to the lack of leading-edge on the rear face of the plate and an approximately speed-independent $h_{\text{front}}$.  Indeed, the obtained force relation $F_{\text{drag}}=K|z|+\lambda\rho A v^2$, which we obtained from experiments as well as continuum modeling, displays only the dynamic inertial correction of DRFT as expected. These results concur with our hypothesis and confirm the DRFT prediction for submerged sideways intrusion. For similar reasons as just discussed, we expect symmetric vertical intrusion of plates to also invoke a negligible structural correction; see Supplementary Information (Figure S3) for details and confirmation against DRFT. \textcolor{black}{Note that in our plate drag studies, we have restricted our intrusion depths to within an $O(1)$ factor of the plate width. This depth range indicates the approximate limits of RFT, as beyond such depths, the assumptions of RFT (such as a linear dependence of granular resistance with depth) begin to degrade \cite{guillard2014lift}.}\par 

\noindent \textit{Four-flap runner:} While the dragged plates are forced to move at set speeds, we also study a self-propelling locomotor, a four-flap runner, whose locomotion speed is determined via the interactions of the locomotor's self actuated limbs (flap motion) and the substrate dynamics (geometric details are in {Table S1 of the Supplementary Information}). The low number of flaps, along with the large flap length to inner radius ratio minimizes the interaction between neighboring flap intrusions of the runner's resultant granular flow. \par 

\begin{figure*}[ht!]
\centering
\includegraphics[trim = 5mm 135mm 20mm 0mm, clip, width=1.0 \linewidth] {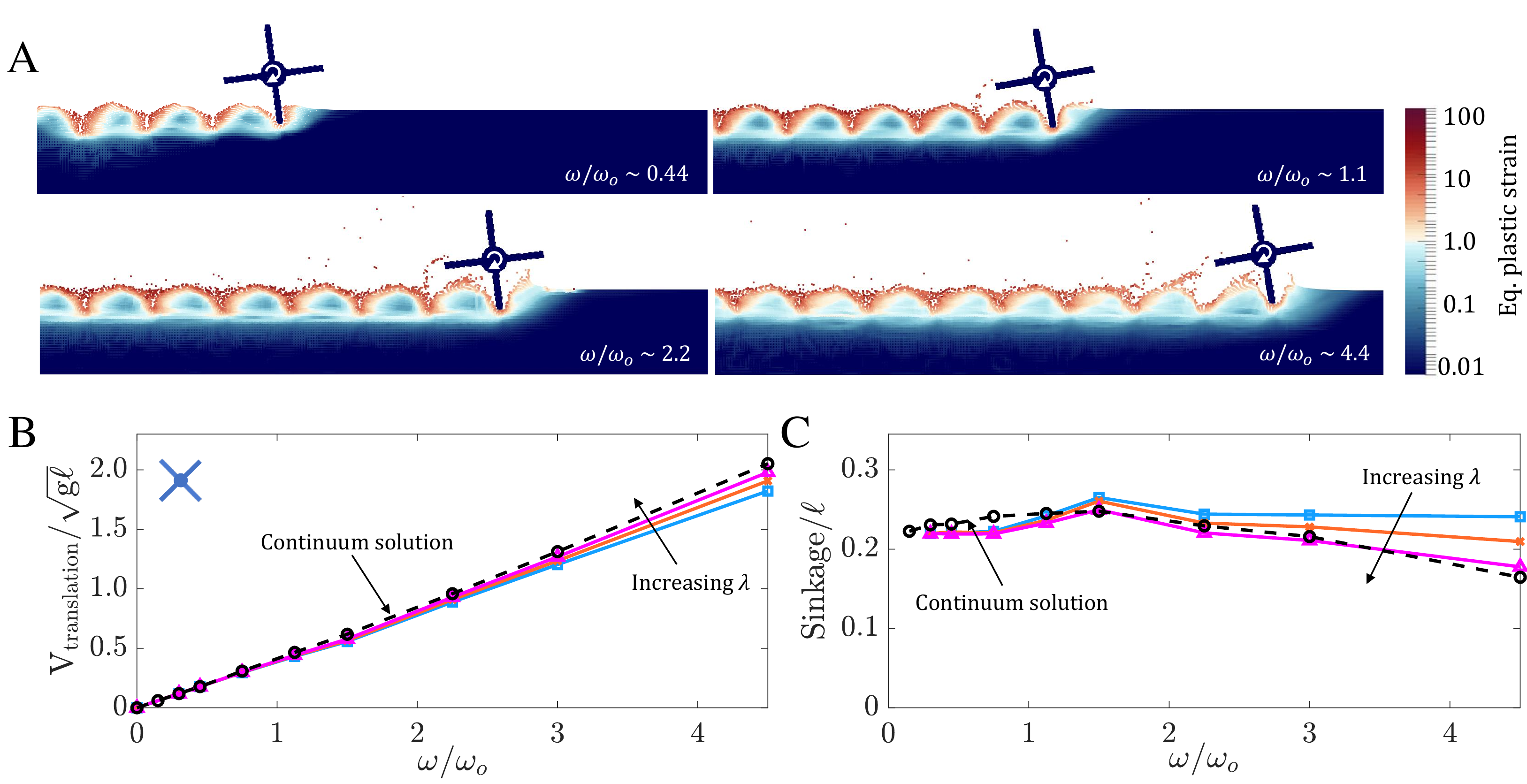}
\caption{\textbf{Running on granular media:} \emph{(A)} Variation of equivalent plastic strain at increasing angular velocities $\omega$ for four-flap runner locomotion ($\omega_o=65$ RPM). See Movie S8 in the Supplementary Information. Continuum solutions from MPM (black dotted line with `o' markers) and DRFT solutions (solid lines) for translational velocity \emph{(B)} and sinkage \emph{(C)} versus angular velocity, $\omega$, in four-flap runner locomotion. DRFT solutions for $\lambda=0,2,4$ pictured. \textcolor{black}{The results in \emph{(B)} and \emph{(C)} are non-dimensionalized as explained in figure \ref{fig:3} with $\ell=190$ mm (runner's outer-diameter).}
}
\label{fig:7}
\end{figure*}

The runner takes inspiration from the experiments of Li et al. \cite{li2013terradynamics} and Zhang et al. \cite{zhang2013ground} with running C-legged robots (similar to Fig \ref{fig:1}C). Li et al. \cite{li2013terradynamics} drove their robots with dimensionless spin ratios ($\omega/\omega_o$) ranging over $0-1.25$ 
\textcolor{black}{$\big((\omega_{max}, \omega_o) = (240,190)$ RPM$\big)$} and observed a decreasing slip with increasing angular velocity in their experiments. Similarly, Zhang et al. \cite{zhang2013ground} tested locomotion over a larger $\omega/\omega_o$ range of $0-3.8$ 
\textcolor{black}{$\big((\omega_{max}, \omega_o) = (720,190)$ RPM$\big)$} and observed that in the higher range of spins, the sinkage in their experiments breaks away from trends observed by Li et al\cite{li2013terradynamics}, i.e. robots elevating above their resting depth. Their running robots display qualitatively opposite behaviors to grousered wheels: as spin increases, runners sink less and move faster whereas wheels sink more and travel slower. We explore if the fundamental physics of such qualitatively reversed behavior is already embedded in our continuum modeling and consequent DRFT framework. Our current continuum modeling capabilities being limited to plane-strain (2D) problems, we cannot implement a full C-legged robot running in 3D. We take the four-flap runner as a representative of the family of runners, which show a decrease in effective-slip and sinkage with increasing angular rotation rates, and explore our 2D continuum model's capability in modeling such behaviors.\par

In the continuum modeling, the dimensionless mass ratio of the runner, given by $m/\rho_c \ell^2W$ for $W$ the out-of-plane width, is set to be in the same range ($\approx$6) as the corresponding twenty-grousered wheels shown previously to keep the comparison between runners and grousered wheels relevant.
 For similar reasons, we keep the runner diameter similar to that of grousered wheel (190  mm vs 212  mm). The angular velocity of the runner is varied over a range of 10 RPM to 300 RPM which corresponds to a dimensionless spin ratio range varying from  $\omega/\omega_o = 0 - 4.5$  ($\omega_o=$ 65 RPM). The continuum results (see Figure \ref{fig:7}B and C) show qualitative agreement with the findings of Li et al. \cite{li2013terradynamics} and Zhang et al. \cite{zhang2013ground} --- with increasing spin rate, a decrease in effective slip and an elevation of the wheel above the rest depth is observed. Incidentally, the turnover in elevation for our runners was found at a spin ratio $\sim1.4$, similar to that obtained by Zhang et al. \cite{zhang2013ground}.\par 

We now use the continuum model as a baseline reference to evaluate the DRFT performance for runners. Figure \ref{fig:7}A shows the variation of the equivalent plastic strain for four different angular velocities in the continuum model. As expected, due to the relatively large separation between intruding legs, there is no visible self-interaction of the granular material between intrusions and the free surface height directly behind intruding legs remains unchanged, which suggests a minimal role of the dynamic structural correction. This observation guides us to model these scenarios using DRFT with typical, $O(1)$, $\lambda$ values ($\lambda = 0,2,4$) and \emph{no} dynamic structural correction. Figure \ref{fig:7}B and C show the resulting steady-state sinkage and translation velocity at various angular velocities from DRFT calculations (solid lines). It seems that DRFT captures the kinematic trends of the reference solution, approaching quantitative accuracy for $\lambda\sim4$. With this result, it is encouraging to note that DRFT has captured the dependence on $\omega$ in both runners and grousered wheels, which behave in opposite ways as $\omega$ increases.\par 

Our four-flap runner study also explains the observations of the above-mentioned C-legged robot studies. We believe the quasistatic RFT modeling in Li et al.  \cite{li2013terradynamics} was sufficient because the dynamic inertial correction was still small in their tested range (in our study, the dynamic inertial correction becomes noticeable only above a $\omega$-ratio of $~\sim1.2$). Zhang et al. \cite{zhang2013ground} go to higher spins, revealing the non-trivial elevation and slip trends due to rate that we see in continuum and DRFT solutions.
\par

\section*{Conclusions}
{\color{black}In this work, we have focused on evaluating the} effectiveness and implications of a continuum model for problems of granular intrusion up to high-speeds, which allows for detailed modeling of complex multiphase inhomogeneous granular systems. \textcolor{black}{We have observed two surprising results. First, a continuum model based only on a constant friction coefficient and tension-free separation is able to model complex granular intrusions well in a variety of scenarios. Second, we find that just two macro-inertial corrections to RFT allows successful modeling of granular intrusions across speed regimes.} 

These results were obtained progressively.  By analyzing the simple continuum model's solutions, an understanding of the key physics involved in such complex intrusion scenarios was identified, which in turn motivated the ingredients of the dynamic resistive force theory (DRFT).  DRFT allows for \textcolor{black}{robust, near real-time modeling} of granular intrusion in a large variety of cases\textcolor{black}{,} including self-propulsion. Our study of rigid intrusion into granular media indicates that the force response upon intrusion consists of two primary rate-dependent modifications: (1) a dynamic inertial correction, and (2) a dynamic structural correction. The dynamic inertial correction accounts for the momentum transfer to the surrounding material, whereas the dynamic structural correction describes how a rapidly moving intruder can change the pressure head by modifying the free surface. Both effects are related to the macro-inertia of the media. For the scenarios considered here, micro-inertial effects \textcolor{black}{(per a $\mu(I)$ rheology)} are not significant even if the motion appears `fast' --- previous work on rapid projectile penetration \cite{dunatunga2017continuum} indicates that the high pressures that develop around rapid intruders keep $I$ relatively small. \textcolor{black}{Hence, the observed rate-dependent dynamics arise under a rate-independent rheology, as we have used, since macro-inertia (stemming from $\rho \dot{v}_i$ in the momentum PDE) alone brings about the observed rate effects. In terms of limitations, it is known that quasistatic RFT loses accuracy when intruders are too deep, as the linear force versus depth dependence eventually plateaus in the lift direction \cite{guillard2014lift} for slow intruders. We expect the same constraints on depth to apply to DRFT as well.}

Dynamic RFT has enough generality to explain two opposing scenarios: weakening of the GM during grousered wheel locomotion, as well as strengthening of the GM during rapid running. We have shown that DRFT accurately predicts the GM system behaviors in the limiting cases, i.e. when one of the two dynamic effects is dominant. Further studies will be required to fully test the model for mixed cases where both dynamic corrections are significant. We have assumed additivity, in line with previous notions of a `static' component and an inertial component of the intrusion force \cite{gravish2014plow, umbanhowar2010granular, katsuragi2007unified,schiebel2019mitigating}.
However, it is possible a more complicated functional combination may arise. 

Although our study mainly focused on dry non-cohesive granular media, the formulation of DRFT in granular flows suggests the existence of other similar reduced-order models in other materials. A combination of experiments and continuum modeling proved vital in this study for verifying the underlying physics. The proposed continuum framework can easily be modified to cater for a large variety of materials once their constitutive equations are known. Future work may explore faster methods of predicting flows, along with various complex intruders to systematically determine the form of the dynamic structural correction. Further studies could also explore the existence of similar reduced-order models for related classes of materials like non-critical state GM, cohesive sands/muds, and fluid-saturated sands. \par 

\section*{Material and Methods}
\subsection*{Experiments (Wheel Locomotion)}
To perform systematic experiments of free-wheel locomotion, we built a simple, automated `terramechanics testbed'. A powerful gear motor (capable of providing up to $70$ RPM at $14.1$ Nm ) is mounted in a carriage (Figure \ref{fig:2}A) which moves freely along vertical and horizontal linear bearings. We control the effective vertical loading of the wheels through a combination of weights and pulleys. The system runs trials in a fluidizing bed of poppy seeds, (a dry non-cohesive granular media) across a bed length of 1 m, allowing for controlled resets of terrain by blowing air up from the bottom. \textcolor{black}{The poppy seeds act as the representative material for the class of non-cohesive granular materials in our study. We specifically choose them due to the ease of running wheel locomotion experiments within them and previous experience using RFT.} This fluidization redistributes the grains evenly into a homogeneous medium after each experiment, giving nearly identical terrain for each test \cite{li2009leggedrobot}. Along with the terrain fluidization, the testbed also has the capability to reset itself: after each run, a linear actuator and a winch work together to drag the wheel carriage back to its starting position. Various system dimensions/specifications are listed in Table S1 of the Supplementary Information.  \par

For experimental visualization of the granular flow around the wheels (Figure \ref{fig:3}C), we also perform PIV analysis of the wheel locomotion at different $\omega$ values. We place the wheel adjacent to the transparent side wall of the poppy seed container and perform the locomotion trials. Images of the flow field are captured with an AOS high-speed camera mounted on a tripod at a resolution of $1280$x$1024$ and a framerate of $500$ FPS. We expect minor variations in the flow fields due to the friction experienced by the material flowing next to the sidewall. The open-source PIVLab package was used in MATLAB for the analysis.

\subsection*{RFT modeling}
To capture the experimental dynamics of the wheel locomotion trials with a reduced-order model, we implement RFT simulations using independent experimental variables and an implicit iterative scheme. A sample simulation diagram is shown in Figure \ref{fig:4}A. Utilizing the rigid wheel assumption, the wheel surfaces are discretized into smaller sub-elements which as a whole approximate the total geometry. The orientation angle ($\beta$), velocity angle ($\gamma$), effective depth from the free surface ($|z|$), and area ($dA$) of each sub-surface is used along with RFT-assumptions of locality and additivity of granular resistive forces, and a `leading edge hypothesis' (discussed earlier) to find net the resistive force and moment. In doing so, equation \ref{eq:RFTformula1} is evaluated using established RFT coefficients from \cite{li2013terradynamics} and the associated scaling coefficients from {Table S1 of the Supplementary Information}. A momentum balance in the x and z coordinates then models wheel motion in the horizontal and vertical direction. The effective heights of wheel grousers were also taken to be one third of their true physical length (based on experimental PIV data) to account for the shadowing effect \cite{suzuki2019study}. Convergence studies of the force response determined the discretization fineness of the wheel shape. Each inner-circumferential subsurface lug was divided into 14 elements and each of the lug surfaces (1 normal and 2 side-wise) was divided into 8 elements. Thus, the wheel had 570 surface elements in total. For the Dynamic RFT implementation, only the effective heights experienced by surface-elements on the rear side of the wheel were modified. This height modification was based on the formulation shown in figure \ref{fig:5}(C). The rear region was taken as the region of rear half of the contact area between sand and wheel (see figure \ref{fig:5}(D)). The division was based on the angle subtended by the contact region at the wheel center.    

\subsection*{Continuum modeling}
We use the material point method (MPM) to carry out the continuum modeling of the system. In MPM, material is discretized as a set of material point tracers that carry the full continuum state. These tracers, representing a chunk of material around their position, are ‘cast’ onto a background simulation grid where equations of motion are solved. Thus, material point tracers act as quadrature points for solving the weak form of the momentum balance equations on a static background simulation grid. A forward-Euler time integration method was used to update the material position and properties. A representative schematic for a MPM time-step update is given in figure \ref{fig:2}C. We model the wheel as a high-stiffness elastic solid with a fixed angular velocity, which is instantaneously enforced on the wheel. In terms of simulation resolution, we use a 200x200 grid representing a domain size of 1m x 1m with initial seeding of 2x2 linear material points per grid cell.




\section*{Acknowledgments}
SA, AK, DG and KK acknowledge support from Army Research Office (ARO) grants {W911NF1510196 and W911NF1810118} and support from the U.S. Army Tank Automotive Research, Development and Engineering Center (TARDEC). The authors declare that they have no competing interests. All data needed to evaluate the conclusions in the paper are present in the paper and/or the Supplementary Materials.

\section*{Supplementary materials}
Materials and Methods\\
Supplementary Text (Section S1 to S3)\\
Figs. S1 to S3\\
Tables S1 \\
Movies S1 to S8 \\
\textcolor{black}{References \textit{(1-38)}}


\clearpage

\section*{Supplementary Information}
\subsection*{S1. Momentum balance approach for granular intrusion}
The essential reason that a velocity-squared additive pressure contribution seems a sensible way to account for macro-inertia can be understood through analysis of the momentum balance equations. In a reference frame moving with an intruder,  balance of momentum at steady state is expressed as
\begin{equation}\label{macromom}\tag{S1}
    0=\rho g_i+\sum_{j=1}^3\frac{\partial}{\partial x_j}\big(\sigma_{ij}(\mathbf{x})-\rho v_i(\mathbf{x})v_j(\mathbf{x})\big).
\end{equation}
Consider a reference case in the quasistatic limit, having slow intruder velocity, $V^{R}_i$, in the lab frame. Let $\sigma_{ij}^R(\mathbf{x})$ and $v^R_i(\mathbf{x})$ represent the corresponding quasistatic stress and flow solutions in the intruder frame.  Thus,
\begin{equation}\tag{S2}
    0=\rho g_i+\sum_{j=1}^3\frac{\partial}{\partial x_j}\sigma^R_{ij}(\mathbf{x}).
\end{equation}
Let us suppose the velocity of the intruder is scaled up to a non-negligible value $V_i=C\, V^R_i$, causing the resultant flow field to exit the quasistatic limit.  Consider the candidate intruder-frame flow field $v_i(\mathbf{x})=Cv^R_i(\mathbf{x})$ and stress field $\sigma_{ij}(\mathbf{x})=\sigma^R_{ij}(\mathbf{x})+\rho (v_i(\mathbf{x}) v_j(\mathbf{x}) - V_iV_j)$.  The proposed flow field assumes similarity of flow between slow and high speed cases, and the stress field properly reduces to the quasistatic solution when $C$ is small. If we suppose the far-field flow vanishes in the lab frame, these candidate fields comprise a valid solution in that they necessarily satisfy equation \ref{macromom}, by design, while preserving the far-field stress condition from $\sigma^R_{ij}$.  Also, kinematic constraints such as incompressibility are guaranteed to transfer from the quasistatic solution to the high-speed solution.

In the intruder frame, the velocity field vanishes at the leading intruder-grain interface, so evaulating the stress at a point $\mathbf{x}^I$ on the intruder interface gives $\sigma_{ij}(\mathbf{x}^I)=\sigma_{ij}^R(\mathbf{x}^I)-\rho V_iV_j$.  It can thus be seen that, under the assumptions made herein, the presence of macro-inertia adds an extra pressure at the intruder-grain interface that goes as density times intruder-velocity squared.  

We emphasize the approximate nature of this analysis.  First, even if $\{\sigma^R_{ij},v^R_i\}$ exactly satisfies the frictional-plastic constitutive relation, the proposed solution $\{\sigma_{ij},v_i\}$ may not. In the main text, we add an additional order-one factor of $\lambda$ ahead of the $\rho v^2$ term to account for deviations caused by this potential mismatch.  Also, importantly, this analysis has assumed the fast flow is a scaled version of the slow flow, which can be an overreaching approximation particularly in cases where dynamic structural corrections modify the granular geometry.

\subsection*{S2. Theoretical derivation of expected linearity between angular and translation velocity from quasistatic RFT}
In the following analysis we present a general proof of linearity between the steady-state translation and the angular velocities for round bodies undergoing free locomotion based on quasistatic RFT. For a rigid body (with fixed mass and shape), if there exists a quasistatic RFT solution with steady-state $(V_{cm},\omega_{cm})$, then there exists another steady-state solution $(\lambda V_{cm},\lambda \omega_{cm})$ for any value of $\lambda$, having the same sinkage, $|z|$.\\
\textbf{Proof:} As per quasistatic RFT, the resistive force per area on any subsurface element of the body can be given as, $\mathbf{t} = \mathbf{t}(\beta, \gamma, z)$,
where, $\beta$ represents the orientation of the subsurface, $\gamma$ represents the local velocity {direction} of the subsurface, and $|z|$ is subsurface depth. Thus, every subsurface experiences a resistive force as a function of its position, orientation, and velocity direction. At a steady state, the sum of these forces in the horizontal direction equals zero, and in the vertical direction, the sum balances the weight of the locomoting body. Any change in the velocity magnitude is not expected to change these forces, since the resistive forces have no dependence on the magnitude of the velocity, only the direction. A multiplicative increase/decrease in the translational and angular velocity pf the wheel by the same constant factor ($\lambda$) will not affect the velocity direction at any subsurface of the body for a given orientation. i.e. $\gamma_{new} =\gamma_{old}$ for any subsurface at a given orientation. Thus, at a given sinkage and orientation, the forces on any subsurface do not change for any value of $\lambda$, leading to no variation in the net force on the locomoting body. Thus, because the forces at any orientation in state $(V_{cm},\omega_{cm})$ result in a steady state, forces at any orientation in state $(\lambda V_{cm},\lambda\omega_{cm})$ will also result in a steady state. 

This analysis holds only when the long-time motion of the locomotor is truly steady --- for example, a non-round wheel has a (wobbly) limit cycle as its long-time behavior but not a steady state.  In the latter case, the analysis still approximately holds if the velocity variations of the wheel do not result in accelerations comparable to $g$. Accelerations of the order of $g$ add a new timescale to the problem which completely changes the problem's dynamics. Similarly, linearity in the locomoting body can break down if the characteristic time scale associated with a flexible body is comparable to the characteristic time scale  $\sqrt{L/g}$. The above hypothesis is also bound to fail if any contributions used in modeling the system are dependent on the velocity magnitude.
{
\subsection*{S3. Additional verification of DRFT in a dynamic inertial correction dominated scenario}

To further verify the robustness of DRFT, we simulate a smaller version of the grousered wheel with continuum modeling as well as DRFT. Doing so also demonstrates the use of scaling analysis in characterizing the free surface profiles for any given class of intruders, using grousered wheels as the test case. The new wheel is similar in shape and composition to the one used in the original study (details in Table S1) but is halved in its spatial dimensions. The granular media is kept the same. The shape and other dimensionless variables $\{\rho_{wheel}/\rho_{c}, \mu_s , h_g/D, N\}$ are kept  same between the two wheels, and hence we assume that the free surface function $\psi$ takes the same form as the reference case for the larger wheel, $\psi = r\omega^2/g$.  With this, we can perform DRFT to predict the dynamics of the small wheel.
\par

\begin{figure*}[ht!]
\includegraphics[trim = 0mm 140mm 10mm 0mm, clip, width=1.0 \linewidth] {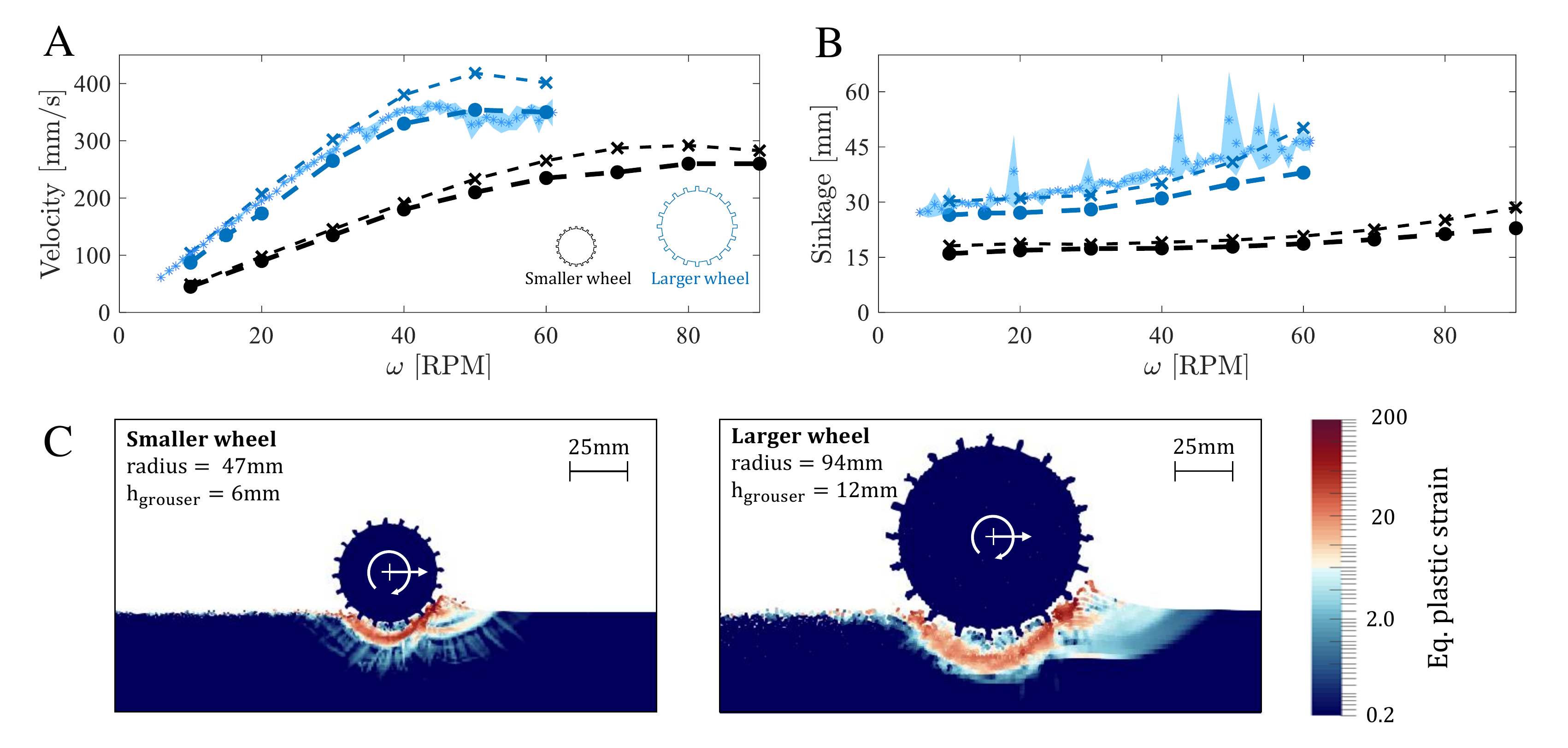}  \par
{Figure S1: \textbf{DRFT performance check for wheels of different dimensions}: Variation  of (A) Translation  velocity, and (B) Sinkage from  experiments (blue data with $*$ markers), continuum  modeling (thin dashed line with filled $\bullet$ markers) 
and DRFT modeling (thick dashed line with $\times$ markers) with respect to angular velocity, $\omega$. Two differently sized wheels are used. The blue color represents the data corresponding to the wheel used in the original study and the black color corresponds to the data from new smaller wheel. (C) and (D) show the spatial variation of equivalent plastic strain rate in smaller and larger wheel cases at $90$ and $60$ RPM resp.}
\label{fig:apdx_fig1}
\end{figure*}

We plot the results from the earlier larger wheel case (in blue) as well as the new scaled down wheel case results (in black) in Figure S1. We use continuum modeling for verifying the accuracy of the results. The new wheel data from DRFT (black dashed lines with '$\times$' markers) matches the reference continuum solution with good accuracy, verifying the robustness of DRFT. }

\begin{table*}[ht!]
\centering
\caption{Wheel dimensions and material properties used in this study}
\begin{tabular}{lr|lr}
\\
\multicolumn{4}{l}{Grousered Wheel Properties $^{\dagger}$} \\ \hline
1. Inner Diameter, $D$ & 188  mm  & 2. Wheel width, $W$ & 140  mm   \\ \hline
3. Grouser height, $h_g$ & 12  mm  & 4. Grouser width, $w_g$  & 6  mm    \\ \hline
5. Gravity, $g$ & 9.8 kg/m$^2$ & 6. Number of grousers, $N$ & 20  \\ \hline
7. Effective horiz. inertia & 23.0 kg & 8. Effective vert. inertia & 6.3 kg \\ \hline        
\\
\multicolumn{4}{l}{ Four-Flap Wheel Properties} \\ \hline
1. Inner Diameter, $D$ & 50  mm  & 2. Wheel width, $W$ & 1 m  \\ \hline
3. Flap length, $h_{flap}$  & 70  mm  & 4. Flap width, $w_{flap}$ & 14  mm    \\ \hline
5. Gravity, $g$  & 9.8 kg/m$^2$ &  6. Mass of wheel, $m$  & 10.5 kg\\ \hline
\\
\multicolumn{4}{l}{Material (Poppy seeds, PS) Properties}\\ \hline
\multicolumn{2}{l}{1. Density, $\rho_{grain}$} & \multicolumn{2}{l}{1100  kg/m$^3$}\\ \hline
\multicolumn{2}{l}{2. Critical packing fraction, $\phi_c$} & \multicolumn{2}{l}{0.58}\\ \hline 
\multicolumn{2}{l}{3. Internal friction, $\mu$ (2D,MPM)$^*$} & \multicolumn{2}{l}{0.56}\\ \hline
\multicolumn{2}{l}{4. Wheel-PS surface friction (2D,MPM)} & \multicolumn{2}{l}{0.35}\\ \hline
\multicolumn{2}{l}{5. RFT scaling coefficient, $\xi$} & \multicolumn{2}{l}{0.35}\\ \hline
\multicolumn{4}{l}{$^{\dagger}$ A constant forward force of 4 N was also applied on the grousered wheel MPM} \\
\multicolumn{4}{l}{\quad simulations to calibrate them with the friction-compensating mechanism used } \\
\multicolumn{4}{l}{\quad in the experiments.} \\
\multicolumn{4}{l}{$^*$ The internal friction ($\mu$) for the PS was obtained by calibrating the sinkage of} \\
\multicolumn{4}{l}{\quad the grousered wheel between MPM simulations and experiments, at low angular}  \\
\multicolumn{4}{l}{\quad  velocity (10 RPM) where system is known to display in quasi-steady character.}  
\label{table:1}
\end{tabular}
\end{table*}

\begin{figure*}[h!]
\includegraphics[trim = 5mm 200mm 70mm 0mm, clip, width=1.0 \linewidth] {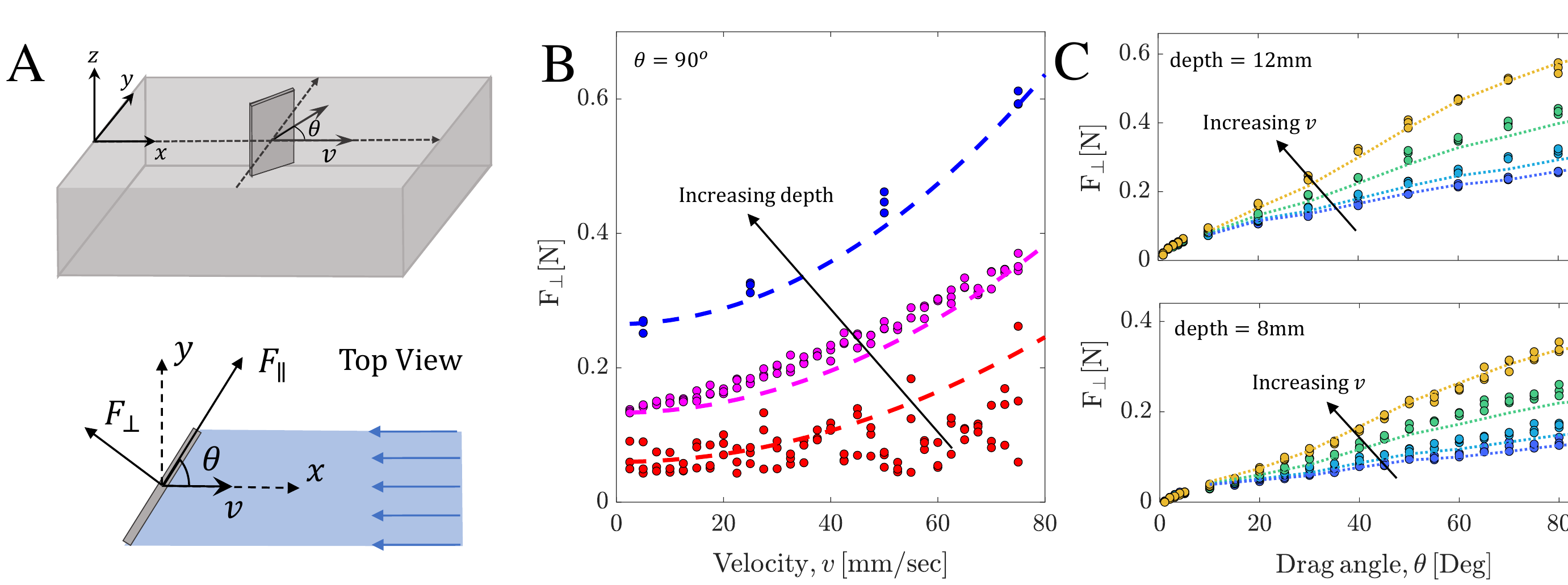} 
 Figure S2: \textbf{DRFT in horizontal intrusion of plates at different oblique angles:} \emph{(A)} Schematic of intrusion scenarios. Experiments (from \cite{schiebel2019mitigating}) use a $\mathrm{30\times 30}$  mm$^2$ thin rigid plate moving with speed $v$ partially-submerged to different depths in a dry granular media of effective bulk density, $\rho$ = $\mathrm{1450 kg/m^3}$ ($\rho_g$ = $\mathrm{2500}$ kg/m$^3$; $\phi$=$0.58$). The plate is rotated to an angle $\theta$ with respect to the velocity direction in the $xy$ (horizontal) plane. \emph{(B)-(C)} Comparisons of experimental measurements of the normal drag component compared to results of the ``static-plus-$\rho v^2$'' drag model, $t_n=t_{\text{static}}(\theta) H(-z)|z|+\lambda \rho v_n^2$, where $t_n$ and $v_n$ are the normal components of the force-per-area and velocity, respectively. We use  $\lambda=1.1$ in all the fittings in this figure and extract $t_{\text{static}}(\theta)$ from the lowest-speed experimental data. \emph{(B)} Experimental data (colored circles) and model results (dashed lines) for normal drag force on plates at a constant angle $\theta =90^o$ for three depths ($\mathrm{[6,8,12]}$  mm, measured from free-surface to plate bottom). Black arrows show the direction of increasing depths.  \emph{(C)} Experimental data (filled circle) and model results (dashed lines) for normal drag force on plates with velocities $\mathrm{[50,250,500,750]}$ mm/s (shown by different colors) at various $\theta$ and two depths ($\mathrm{[8, 12]}$ mm). Black arrows show the direction of increasing velocities at each depth.  
\label{fig:apdx_fig2}
\end{figure*}

\begin{figure*}[ht!]
\includegraphics[trim = 20mm 65mm 40mm 0mm, clip, width=0.9 \linewidth] {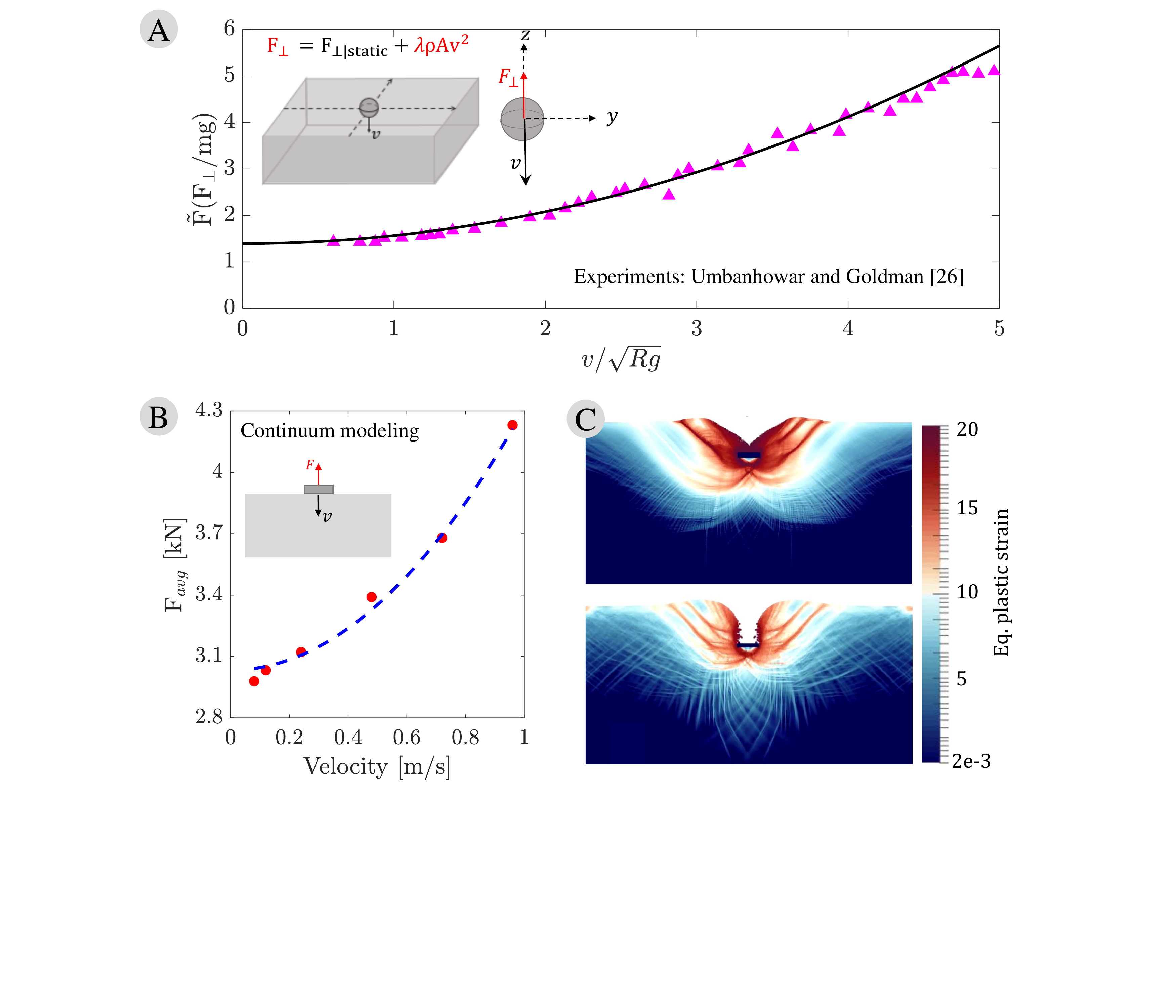}  \par
Figure S3: \textbf{DRFT in downward intrusion:} \emph{(A)} Schematic and normalised force variation for vertical intrusion based on Umbanhowar and Goldman \cite{umbanhowar2010granular} sphere intrusion data at critical state packing fraction {(at a fixed depth). Umbanhowar and Goldman generated this data by measuring the acceleration of spheres impacting the granular beds at fixed depths being impacted at various different initial velocities. Triangles (pink) represent experimental data and solid line (black) represents the $\lambda \rho v^2$  fit to the data. We find $\lambda = 1.4$ to be an appropriate value. \emph{(B)} Data (red circles) and \emph{(C)} visualization of equivalent 2D plane-strain continuum simulations for vertical plate intrusions at low-speed (top) and high-speed (bottom). The continuum data in \emph{(B)} can be qualitatively compared to experiments in \emph{(A)} but not quantitatively because the two utilize different intruder shapes and out-of-plane conditions. For \emph{(B)}, $0.16$ m plates were intruded in a granular volume of density $2500$ kg$/$m$^3$ and internal friction $\mu=0.62$; $\lambda=2.8$ was used for DRFT fittings (blue dotted lines). } The MPM plots in \emph{(C)}  visualize the material flow field around the intruding body. These results indicate the dominance of the dynamic inertial correction in the drag forces. The reason is that similar to horizontal plate intrusions, which in the plane-strain approximation often have only one bulk flow region (creating a Coulomb-wedge structure \cite{gravish2014plow}), the vertical plate intrusions also create two non-interacting bulk flow regions to either side. The free surface regions where flow trajectories eminating from the plate bottom terminate remain at the same height on both sides, in both the low and high-speed intrusion cases. Thus, the resistive force response encounters minimal contribution from dynamic structural correction; the dynamic inertial correction dominates.
\label{fig:apdx_fig3}
\end{figure*}

\begin{figure*}[ht!]
\includegraphics[trim = 0mm 0mm 0mm 0mm, clip, width=1.0 \linewidth] {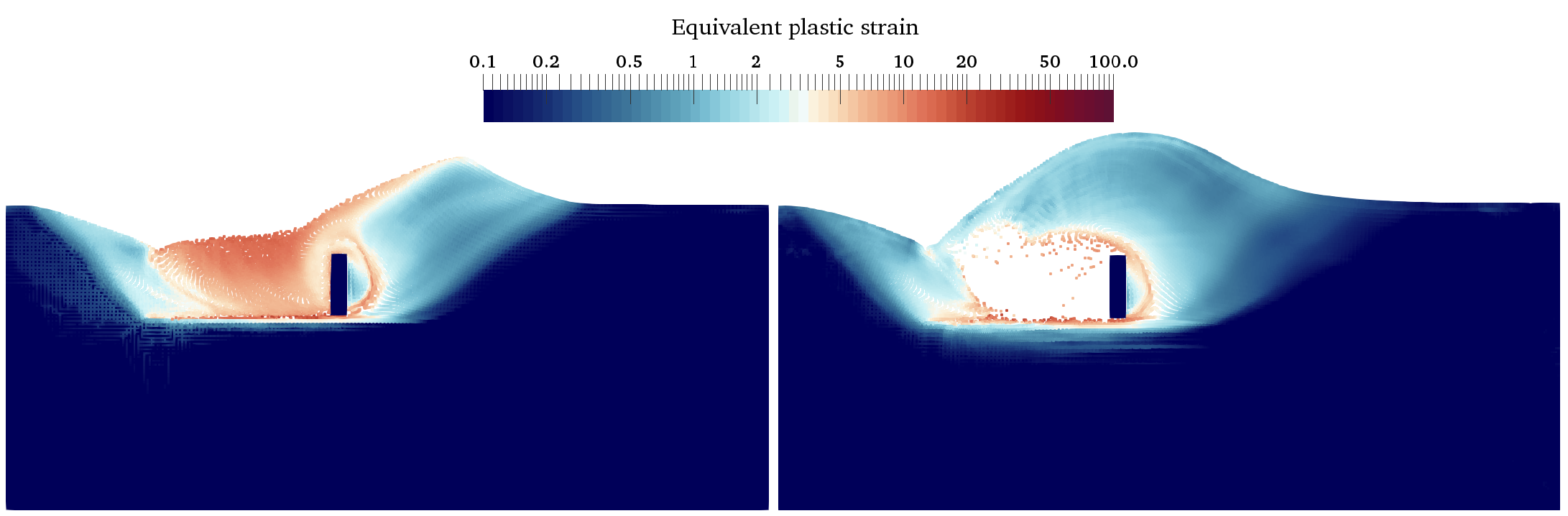} \par
Movie S1: 
{Variation of equivalent plastic strain in 2D plane-strain continuum simulations for horizontal plate intrusion at both low (0.04 m/s, left) and high speeds (0.64 m/s, right). The videos show that the material flow profile (a Coulomb-wedge structure) ahead of the intruding plates, where most of the drag originates, does not change between low and high speed intrusions. As a result, the free surface height, the material flow geometry, and the shape around the intruders remains similar regardless of intrusion rate. Thus, the resistive force response encounters minimal contribution from the dynamic structural correction and the speed dependence is dominated by the dynamic inertial correction.}
\label{fig:mov1}
\end{figure*}

\begin{figure*}[ht!]
\includegraphics[trim = 0mm 0mm 0mm 0mm, clip, width=1.0 \linewidth] {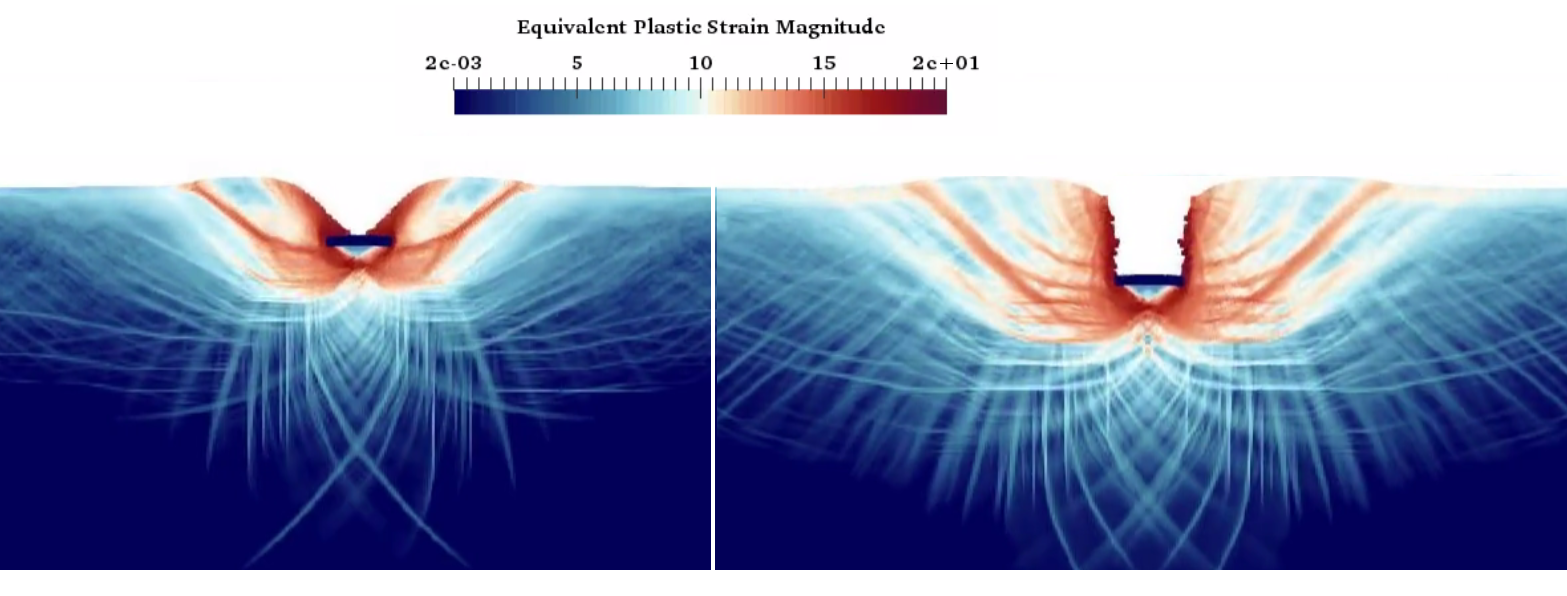} \par
Movie S2:
{Variation of equivalent plastic strain in 2D plane-strain continuum simulations for vertical plate intrusion at low-speed (left) and high-speed (right). The videos show that similar to horizontal intrusions, two independent Coulomb-wedge type structures are formed during the vertical intrusion of plates. The free surface height, the material flow geometry, and the shape around the intruders remain similar regardless of intrusion rate. Thus, the resistive force response encounters minimal contribution from the dynamic structural correction, but has a significant dynamic inertial correction. }
\label{fig:mov2}
\end{figure*}

\begin{figure*}[ht!]
\includegraphics[trim = -30mm 0mm 0mm 0mm, clip, width=0.9 \linewidth] {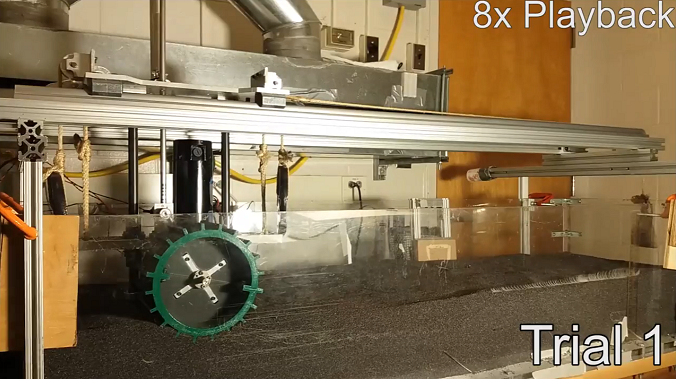} \par
Movie S3: 
{Sample experimental runs of free wheel locomotion: Lateral motion of the wheel is restricted with a double rail and bearing system. The granular material is reset to a loose packing state at the beginning of each trial by air fluidization of the granular bed. Photo Credit: Andras Karsai, Crab Lab, Georgia Institute of Technology. }
\label{fig:mov3}
\end{figure*}

\begin{figure*}[ht!]
\includegraphics[trim = 0mm 0mm 0mm 0mm, clip, width=1.0 \linewidth] {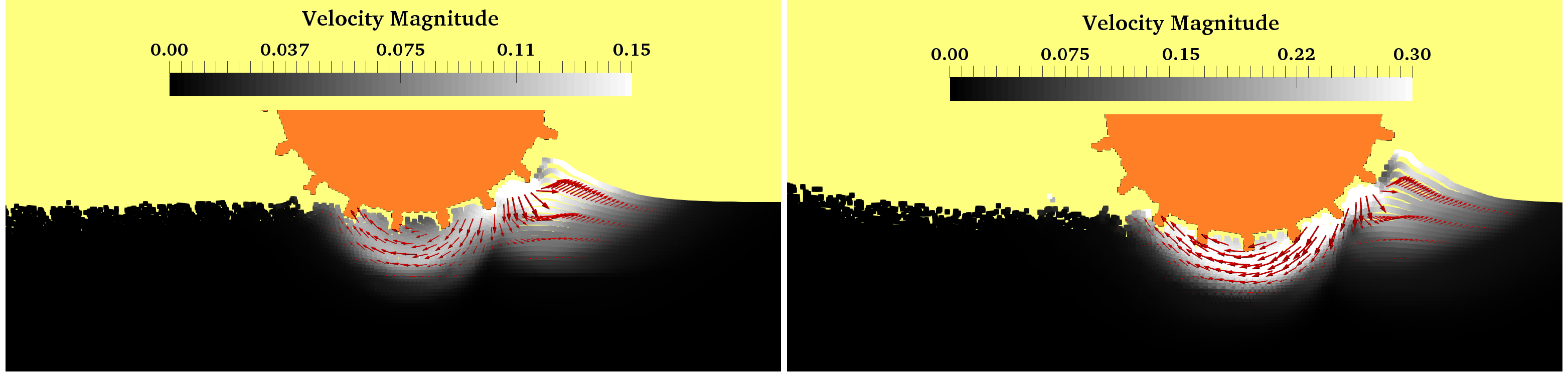} \par
Movie S4: 
{Visualization of material flow using continuum modeling at 30 RPM and 60 RPM wheel locomotion: The color bars indicate the variation of velocity magnitude in the granular media domain. The material flow clearly indicates an increase in the size of the flow zone with increasing rotation rates.}
\label{fig:mov4}
\end{figure*}

\begin{figure}[ht!]
\includegraphics[trim = 0mm 0mm 0mm 0mm, clip, width=0.45 \linewidth] {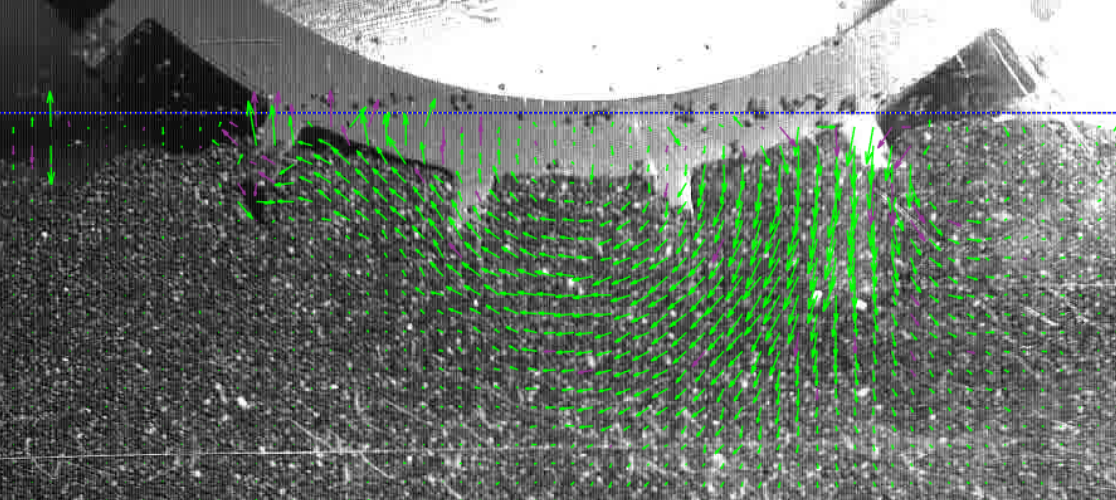}
\qquad \qquad
\includegraphics[trim = 0mm 0mm 0mm 0mm, clip, width=0.45 \linewidth] {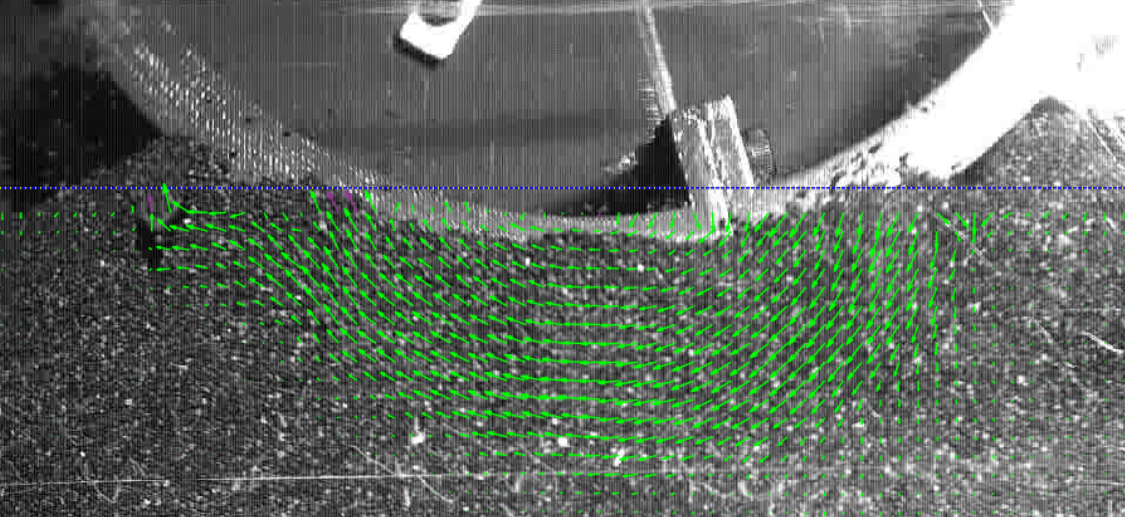} \par
Movie S5 and S6: 
{A PIV visualization of 30 RPM and 60 RPM wheel locomotion experiments. Increasing flow zone with increasing rotation rates similar to continuum modelling (Movie S4) were observed. Photo Credit: Andras Karsai, Crab Lab, Georgia Institute of Technology.}
\label{fig:mov5and6}
\end{figure}

\begin{figure*}[ht!]
\includegraphics[trim = -40mm 0mm 0mm 0mm, clip, width=0.9 \linewidth] {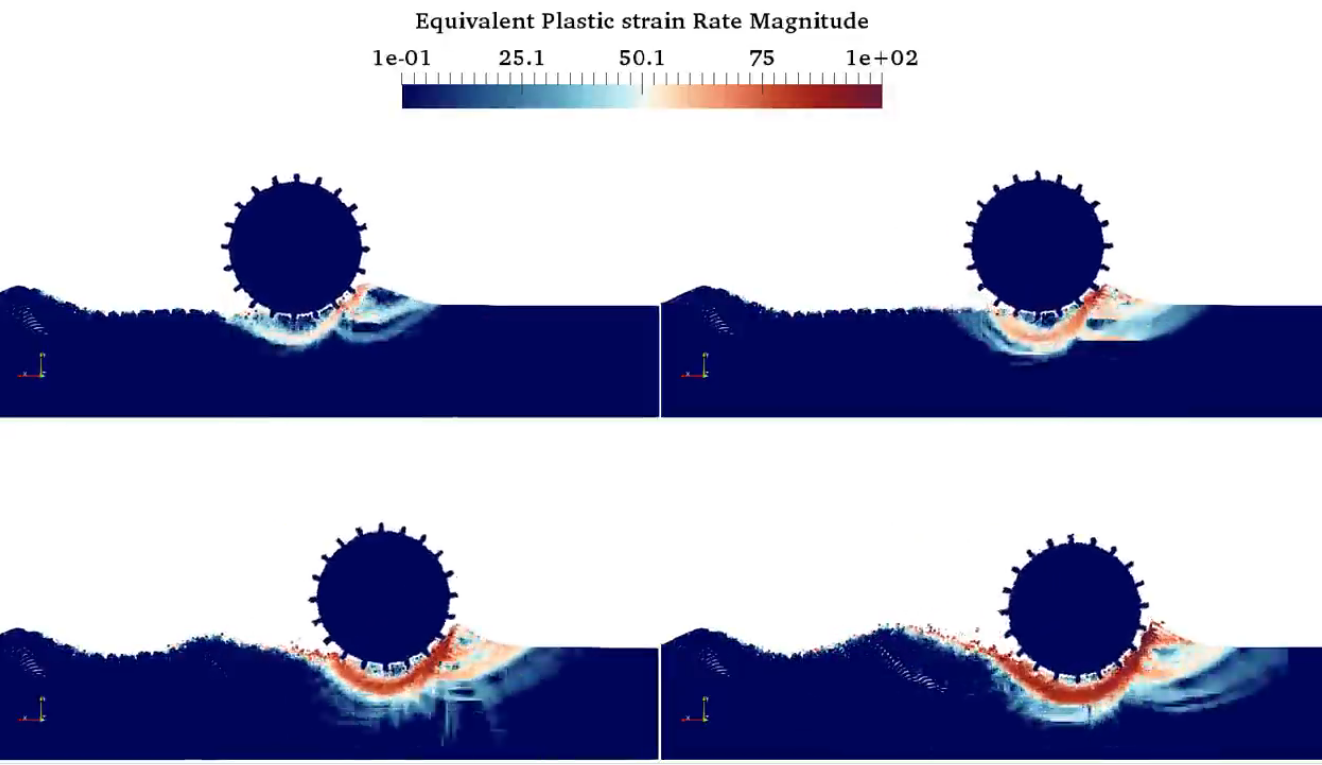} \par
Movie S7: 
{Visualization of material flow during wheel locomotion at [10,30,60,90] RPM. We plot equivalent plastic strain rate from continuum modeling, a scalar measure of the size of plastic strain. For a better visualization, all the wheels start at the same rotation rate of 30 RPM in the beginning of the simulations and switch to [10,30,60,90] RPM after a fixed finite horizontal motion, which is synchronized to be the beginning of the video.   
}\label{fig:mov7}
\end{figure*}

\begin{figure*}[ht!]
\includegraphics[trim = -40mm 0mm 0mm 0mm, clip, width=0.9 \linewidth] {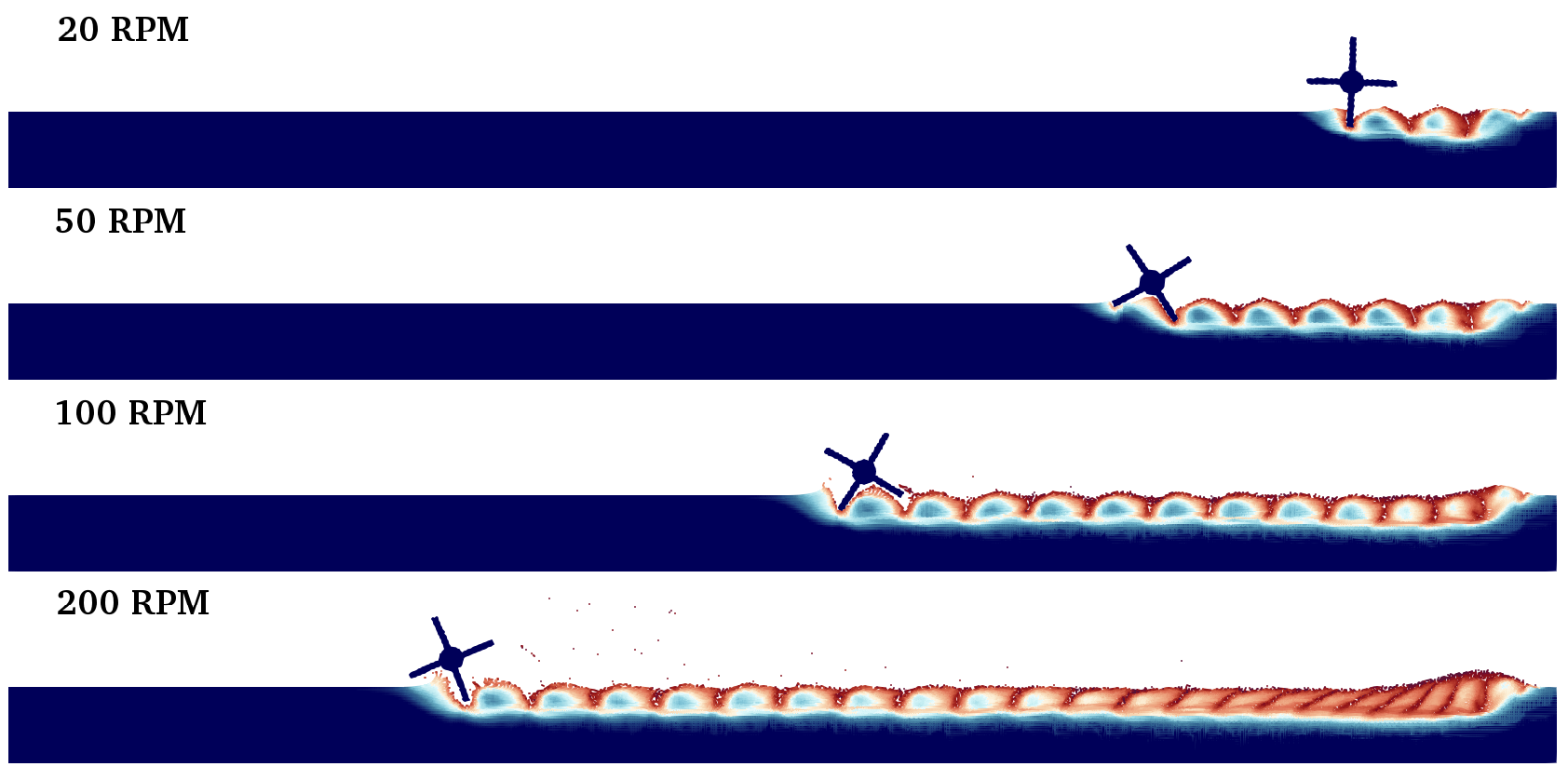} \par
Movie S8: 
{Visualization of material flow during 4-flap runner locomotion at [20,50,100,200] RPM. We plot equivalent plastic strain from the continuum model, a scalar measure of the size of plastic strain. As expected, the flowing regions are separated by flap intrusions and thus do not interact enough to cause an effective free surface change for the material trajectories. Thus, only a dynamic inertial update was used in DRFT modeling of this case. 
}\label{fig:mov8}
\end{figure*}

\end{document}